\documentclass[12pt]{article}
\usepackage[dvips]{graphicx}

\usepackage{graphics,epsfig}

\usepackage{psfrag}

\renewcommand{\bar}[1]{\overline{#1}}

\renewcommand{\bar}[1]{\overline{#1}}

\def\ru1{\rule[-0.4truecm]{0mm}{1truecm}}

\begin{document}

\rightline{\small KCL-PH-TH/2012-11, LCTS/2012-06, CERN-PH-TH/2012-058}
\vspace{10mm}

\centerline{\Large \bf Does the `Higgs' have Spin Zero?}

\vspace{7mm}

\centerline{{\bf John Ellis$^{1}$} and {\bf Dae Sung Hwang$^{2}$}}

\vspace{5mm}

\vspace{4mm} \centerline{\it $^1$Theoretical Particle Physics and Cosmology Group, Physics Department,}
\centerline{\it King's College London, Strand, London, UK;}
\centerline{\it Theory Division, Physics Department, CERN, 1211 Geneva 23, Switzerland}

\vspace{4mm} \centerline{\it $^2$Department of Physics, Sejong
University, Seoul 143--747, South Korea}

\vspace{20mm}

\centerline{\bf Abstract}

\vspace{10mm}

\noindent
The Higgs boson is predicted to have spin zero. 
The ATLAS and CMS experiments have recently reported of an excess of events
with mass $\sim 125$~GeV that has some of the characteristics expected
for a Higgs boson. We address the questions whether there is already any evidence
that this excess has spin zero, and how this possibility could be confirmed in the near future. 
The excess observed in the $\gamma \gamma$ final state could not have spin one, leaving
zero and two as open possibilities. We review the angular distribution of $\gamma \gamma$
pairs from the decays of a graviton-like spin-two boson produced in gluon-gluon collisions, which
is well-defined and distinct from the spin-zero case. We also calculate the distributions for lepton pairs
that would be produced in the $W W^*$ decays of a spin-two boson, which are very different from
those in Higgs decays, and note that the kinematics of the event selection 
currently used in the analysis of
the $W W^*$ final state have reduced efficiency for spin two.

\vspace{0.5cm}

\noindent \vspace*{12mm}


\vspace{3mm}

\noindent  {\small {\bf PACS numbers:} 14.70.Kv Gravitons, 14.80.Bn Standard-model Higgs bosons,
13.85.Qk        Inclusive production with identified leptons, photons, or other nonhadronic particles,
13.88.+e Polarization in interactions and scattering}




\newpage


\section{Introduction}

The Higgs boson is predicted to have spin zero. Since all known elementary
particles have non-zero spin, this is a crucial property to be checked by
experiment before one could claim that the quest for this `Holy Grail' of
particle physics has been concluded successfully. Reflecting the importance
of this issue, there have been many
studies of the potential of the LHC experiments for measuring the spin of any
candidate for the Higgs boson~\cite{1,2,3,4,5,6,7,8,9,9a,10,11,12,13}. 

Most of these
papers proposed to look at spin correlations in $ZZ$ or $Z Z^*$ decays
using four-charged-lepton final states ~\cite{2,4,5,7,9,9a,10,11,12,13}.
To our knowledge, the only published study of a spin-two state $X$ decaying into $\gamma \gamma$ has been~\cite{10},
see Appendix A. 
Ref.~\cite{6} studied the production of $X \to W^+ W^- \to \ell^+ \ell^- \nu {\bar \nu}$
final states via vector-boson fusion,
distributions for the transverse angles of charged leptons in particles $X \to W^+ W^- \to \ell^+ \ell^- \nu {\bar \nu}$
decays were considered in~\cite{7}, and high-mass $X \to W^+ W^- \to \ell^+ \ell^- \nu {\bar \nu}$ decays were considered in~\cite{13}.
However, in all these papers only the cases where $X$ has spin 0 or 1 were considered. Ref.~\cite{10}
discussed spin-2 decays into $W^+ W^-$ but did not discus in detail charged-lepton angular distributions. Refs.~\cite{1,8}
considered production in $e^+ e^-$ collisions, in association with $Z$ and ${\bar t} t$, respectively.

The ATLAS~\cite{ATLASH} and CMS~\cite{CMSH} collaborations have recently reported evidence for
excesses in $\gamma \gamma$ and $Z Z^*$ that are consistent with
expectations for a Standard Model Higgs boson, an interpretation
supported by broader enhancements of less significance in $WW^*$, 
$\tau \tau$ and ${\bar b} b$ final states. The statistics in $Z Z^*$ decays
are as yet insufficient for an attempt to constrain the `Higgs' spin, so in this
paper we consider other ways to obtain an indication what it may be.

A spin-one state cannot decay into two identical vector bosons, so
a peak observed in the $\gamma \gamma$ final state must have spin zero
or two~\footnote{In principle, one could consider also higher even spins, but
these would entail production and decay mechanisms involving orbital
angular momentum factors that we ignore here.}. Fermion-antifermion final
states could come from spin zero or spin one~\footnote{Again neglecting
orbital angular momentum.}, so observation of a `Higgs' signal in either
of the $\tau \tau$ or ${\bar b} b$ final states would favour the spin-zero
hypothesis over the spin-two option. However, so far only CMS reports
any enhancements in these channels, and they are each $\le 1 \sigma$
for a mass of 125~GeV~\cite{CMSH}, so not conclusive at the present time.
Accordingly, we consider here the $\gamma \gamma$ and $W W^* \to \ell^+ \ell^- \nu {\bar \nu}$ 
final states, which have been observed with greater significance by 
both ATLAS and CMS.

Under the assumption that P-wave fermion-antifermion collisions
can be neglected~\footnote{This is not necessarily the case in string models~\cite{AGT}.}, 
a spin-two particle could be produced either by gluon-gluon
collisions or by vector-boson fusion. We consider here the production of a 
hypothetical spin-two particle $X_2$ via gluon-gluon fusion, which
is the dominant production mechanism for producing a Higgs boson weighing
$\sim 125$~GeV.  For definiteness, we assume that the $X_2$ couplings
are of the same form as a massive Kaluza-Klein graviton~\cite{Giudice,Han,Gleisberg}
and in a string model~\cite{AGT}, 
though without committing ourselves to either framework~\footnote{It was shown 
in~\cite{FGLS} that this form is unique if the $X_2$ couplings to pairs of
vector bosons are of dimension five (the lowest possible) and one assumes gauge and CP
invariance.}.

We first review the angular distribution for
$gg \to X_2 \to \gamma \gamma$ (see Appendix A of~\cite{10}), 
recalling that if graviton-like couplings are assumed the final-state angular distribution
in the $X_2$ centre of mass system is completely determined. 
It is suppressed at large angles relative to the beams, and hence is in principle distinct from the 
isotropic distribution predicted for spin-zero Higgs decay.

We then turn to the $X_2 \to W W^* \to \ell^+ \ell^- \nu {\bar \nu}$ final state, again assuming production by gluon-gluon
fusion  and the same couplings as in massive graviton models~\cite{Giudice,Han,Gleisberg}. 
We note that the ATLAS and CMS searches for $W W^* \to \ell^+ \ell^- \nu {\bar \nu}$ 
final states~\cite{ATLASH,CMSH} already incorporate a hypothesis about the spin of the `Higgs'
candidate. They make use of the observation in~\cite{DD} that a spin-zero
particle decaying into $W W$ (or $W W^*$) would yield final states in which
the $W^+$ and $W^-$ would have opposite polarizations. Since the $W^-$
decays exclusively into {\it left}-handed leptons, whereas the $W^+$ decays
exclusively into {\it right}-handed leptons, the {\it anticorrelation} between the
$W^\pm$ polarizations expected in spin-zero Higgs decay would be transferred
into a {\it correlation} between the momenta of the charged leptons in their decays. 
This correlation would manifest itself in  the distributions of relative $\ell^\pm$ polar 
angles and a preference for a small azimuthal
angle between the $\ell^+ \ell^-$ pair, $\phi_{\ell^+ \ell^-}$, with a relatively small 
invariant mass, $m_{\ell^+ \ell^-}$. Both ATLAS and CMS select events with
 cuts based on these observations~\cite{ATLASH,CMSH}.

We study the types of $\ell^+ \ell^-$ correlations to be expected in the 
$W W^* \to \ell^+ \ell^- \nu {\bar \nu}$ decays of a spin-two state. We find that
their momenta tend to be {\it anticorrelated},  with distinctive features
in both polar and azimuthal angle distributions,
and hence quite distinct from those expected for the decays of a spin-zero state,
Hence, the observation (or not) of Higgs-like $\ell^+ \ell^-$ correlations in
$W W^*$ final states could help provide evidence that ATLAS and
CMS may be observing a spin-zero (-two) state. This possibility should be pursued
with experimental simulations of spin-two $W W^* \to \ell^+ \ell^- \nu {\bar \nu}$ 
decays using the results presented here, which would indicate how much data
would be needed to confirm the result with a significant degree of confidence.

\section{Preliminaries}

\subsection{Production Kinematics}

Ideally, one would prefer to perform such a `Higgs' spin analysis in the most
model-independent way possible. However, the density matrix of a massive spin-two
particle has many parameters, and the available statistics limit the complexity of the
hypotheses one can test currently, so we are led to make motivated simplifying
assumptions about the possible production mechanism of a massive spin-two
state. Bosons are generally produced in $pp$ collisions by ${\bar q} q$, $gg$
or $WW/ZZ$ collisions. However, neglecting orbital angular momentum,
${\bar q} q$ collisions can produce only spin-zero or -one states, 
so we are left with $gg$ and $WW/ZZ$ collisions. Since $gg$ collisions are much
more copious and simpler to analyze, we focus on them.

Neglecting initial-state transverse momentum and radiation, we may regard
the gluons as massless spin-one particles whose momenta are aligned with the
collision axis. As such, if one quantizes angular momentum along this axis,
they are equally likely to be in the helicity states $| 1, \pm 1 \rangle$. We assume that
there is no coherence between the final states in which different gluon helicity
states collide. Therefore the initial-state combinations $| 1, + 1 \rangle | 1, + 1 \rangle, 
| 1, + 1 \rangle | 1, -1 \rangle, | 1, - 1 \rangle | 1, + 1 \rangle$ and $| 1, - 1 \rangle | 1, - 1 \rangle$
are equally likely. 
Accordingly, the $gg$ initial states are a combination of the
$| 2, + 2 \rangle, | 2, -2 \rangle$ and $| 2, 0 \rangle$ polarization states, described by a spin-two
density matrix $\rho_2$ that has only diagonal entries with relative
weights
\begin{equation}
{\rho}_i = {3\over 7} \Big( |22 \rangle \langle 22| + {{1\over 3}}|20 \rangle \langle 20| + |2-2 \rangle \langle 2-2| \Big) \ ,
\label{rhoi}
\end{equation}
where the relative normalization of the $J_z = 0$ component is determined by the
Clebsch-Gordan coefficients $\langle 2, 0 || 1, \pm 1 \rangle |1, \mp 1 \rangle = 1/\sqrt{6}$.

We explore in the following sections the consequences of this observation for
the possible decays of a hypothetical spin-two particle $X_2$ into $\gamma \gamma$
and $W^+ W^-$ final states at the LHC. The $|20 \rangle \langle 20|$ component in the
density matrix does not contribute if graviton-like couplings are assumed~\cite{AGT},
as done here.

\subsection{Polarization States}

Before discussing further the kinematics and dynamics of $X_2$ production and decay, we briefly review and
establish our notation for the polarization states of the spin-one and -two particles
appearing in our analysis.

A massive spin-one particle with momentum $p^{\mu}=(p^0,p^1,p^2,p^3)=(E,0,0,p)$ has
three independent polarization states given by
\begin{eqnarray}
\epsilon^{+\mu}&=&(0, -{1\over {\sqrt{2}}}, -{i\over {\sqrt{2}}}, 0) \ ,
\label{a1}\\
\epsilon^{-\mu}&=&(0, +{1\over {\sqrt{2}}}, -{i\over {\sqrt{2}}}, 0) \ ,
\label{a2}\\
\epsilon^{0\mu}&=&({p\over m}, 0, 0, {E\over m}) \ .
\label{a3}
\end{eqnarray}
If we work in the Lorentz frame where the vector particle is at rest, so that
$p^{\mu}=(p^0,p^1,p^2,p^3)=(m,0,0,0)$,
the three polarization vectors are given by (\ref{a1}), (\ref{a2}),
and for (\ref{a3})
\begin{equation}
\epsilon^{0\mu}\ =\ (0,0,0,1) \ .
\label{a0}
\end{equation}
The polarization vectors $\epsilon^{+\mu}$, $\epsilon^{-\mu}$ and $\epsilon^{0\mu}$
correspond to the quantum states $|1,+1 \rangle$, $|1,-1 \rangle$ and $|1,0 \rangle$, respectively,
with the $z$-axis as the quantization axis.

We now consider the spin states of a spin-two particle $X_2$ with mass $m$, in its rest frame.
The polarizations of $X_2$ can be represented by the following polarization tensors:
\begin{eqnarray}
\epsilon^{s\,\mu\nu}&=&
\Big( |2+2\rangle,\ |2,+1\rangle,\ |2,0\rangle,\ |2,-1\rangle,\ |2,-2\rangle\Big)
\nonumber\\
&=&\Big( \epsilon^{+2\,\mu\nu},\ \epsilon^{+1\,\mu\nu},\ \epsilon^{0\,\mu\nu},\ 
\epsilon^{-1\,\mu\nu},\ \epsilon^{-2\,\mu\nu}\Big)
\nonumber\\
&=&\Big( \epsilon^{+\mu}\epsilon^{+\nu},\ 
{1\over {\sqrt{2}}}(\epsilon^{+\mu}\epsilon^{0\nu}+\epsilon^{0\mu}\epsilon^{+\nu}),\
\nonumber\\
&& 
{1\over {\sqrt{6}}}(\epsilon^{+\mu}\epsilon^{-\nu}+\epsilon^{-\mu}\epsilon^{+\nu}+
2\epsilon^{0\mu}\epsilon^{0\nu}),\
{1\over {\sqrt{2}}}(\epsilon^{-\mu}\epsilon^{0\nu}+\epsilon^{0\mu}\epsilon^{-\nu}),\
\epsilon^{-\mu}\epsilon^{-\nu}\Big)\ ,
\label{b1}
\end{eqnarray}
where $\epsilon^{+\mu}$, $\epsilon^{-\mu}$ are given in (\ref{a1}) and (\ref{a2}),
and $\epsilon^{0\mu}$ is given in (\ref{a0}), since we work in the $X_2$ rest frame.

The polarization tensors given in (\ref{b1}) satisfy the following relations:
\begin{equation}
(\epsilon^{s})^{\,\ \mu}_{\mu}=0\ ,\ \ \ \
p_{\mu}\epsilon^{s\,\mu\nu}=0\ ,\ \ \ \
\epsilon^{s\,\mu\nu}\epsilon^{s'\ *}_{\,\mu\nu}={\delta}^{ss'}\ ,
\label{b2}
\end{equation}
and
\begin{equation}
\sum_{s=-2}^{+2}\epsilon^{s}_{\,\mu\nu}\epsilon^{s*}_{\,\alpha\beta}=B_{\mu\nu\ \alpha\beta}\ ,
\label{b3}
\end{equation}
where
\begin{eqnarray}
B_{\mu\nu\ \alpha\beta}&=&
\Big( {\eta}_{\mu\alpha}-{p_{\mu}p_{\alpha}\over m^2}\Big)
\Big( {\eta}_{\nu\beta}-{p_{\nu}p_{\beta}\over m^2}\Big)\ +\
\Big( {\eta}_{\mu\beta}-{p_{\mu}p_{\beta}\over m^2}\Big)
\Big( {\eta}_{\nu\alpha}-{p_{\nu}p_{\alpha}\over m^2}\Big)
\nonumber\\
&&-\
{2\over 3}\, \Big( {\eta}_{\mu\nu}-{p_{\mu}p_{\nu}\over m^2}\Big)
\Big( {\eta}_{\alpha\beta}-{p_{\alpha}p_{\beta}\over m^2}\Big)\ .
\label{b4}
\end{eqnarray}
We further note for reference that the propagator of the spin-two massive $X_2$ particle is given by
\cite{Giudice,Han,Gleisberg}
\begin{equation}
i\Delta_{\mu\nu\ \alpha\beta}=
{i\, B_{\mu\nu\ \alpha\beta} \over p^2-m^2+i\varepsilon}\ ,
\label{b5}
\end{equation}
though the denominator of this formula is not used in this paper, since we consider only on-shell resonant production of $X_2$.

\section{The Process $gg \to X_2 \to \gamma \gamma$}

\subsection{Preliminaries}

\begin{figure}
\centering
\begin{minipage}[t]{8.0cm}
\centering
\includegraphics[width=\textwidth]{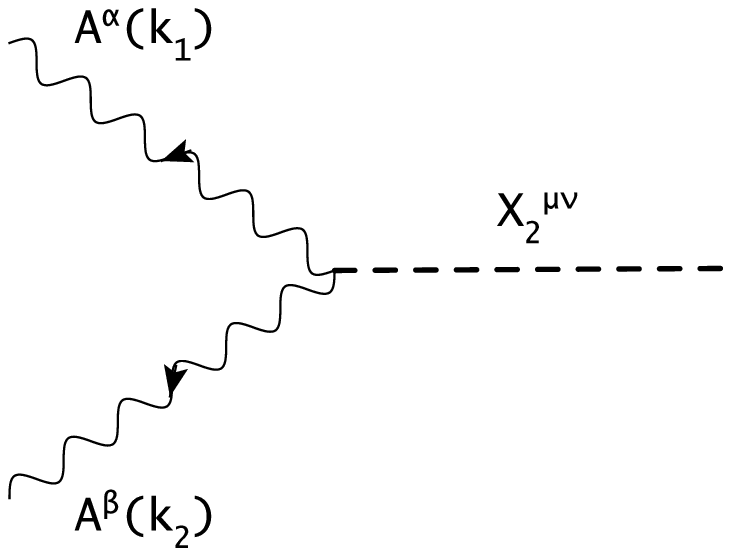}
(a)
\end{minipage}\hspace{1.0cm}
\begin{minipage}[t]{9.5cm}
\centering
\includegraphics[width=\textwidth]{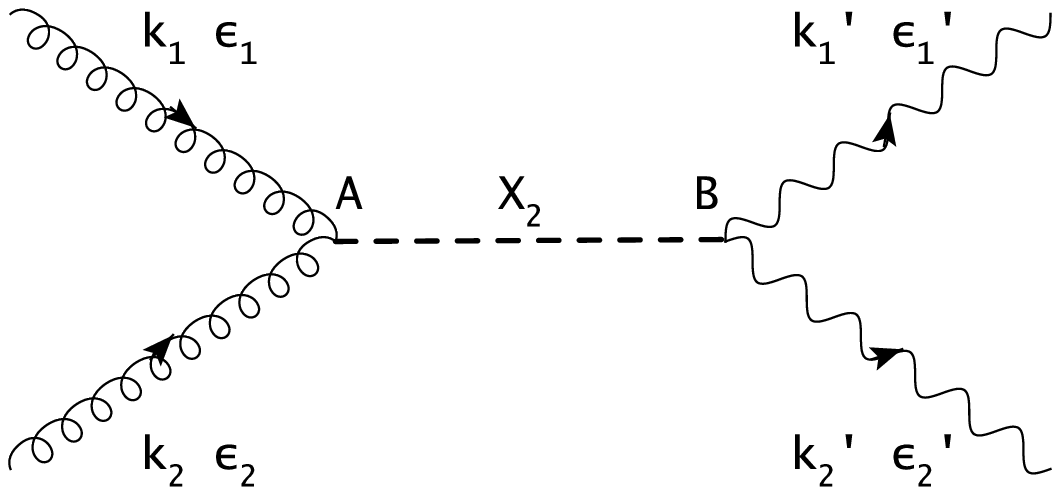}
(b)
\end{minipage}\hspace{1.0cm}
\begin{minipage}[t]{9.5cm}
\centering
\includegraphics[width=\textwidth]{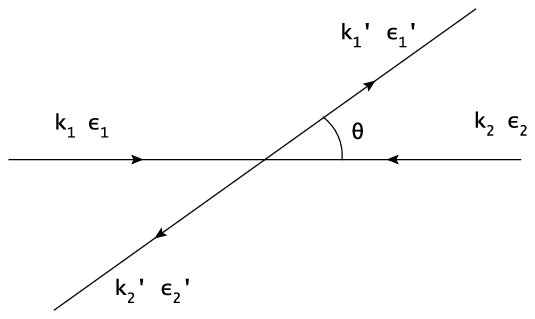}
(c)
\end{minipage}\hspace{0.0cm}
\parbox{0.95\textwidth}{\caption{
\it (a) The vertex coupling $X_2$ to two gauge fields, (b) Feynman
diagram and (c) the kinematics for the process $gg \to X_2 \to \gamma \gamma$.
\label{diagrams}}}
\end{figure}

The three-point vertex for $X_2\gamma\gamma$ or $gg$
is illustrated in Fig. \ref{diagrams}(a), the process $gg \to X_2 \to \gamma\gamma$
is illustrated in Fig. \ref{diagrams}(b), and our notation for the kinematics is
illustrated in Fig. \ref{diagrams}(c). For definiteness,we use the following Feynman riule for
the $X_2\gamma\gamma$ vertex, which was derived in~\cite{Giudice,Han} for the
coupling of a massive Kaluza-Klein graviton:
\begin{equation}
-\, {i\over {M}}\, \Big( W^{(\gamma)}_{\mu\nu\ \alpha\beta}\ +\
W^{(\gamma)}_{\nu\mu\ \alpha\beta}\Big)\ ,
\label{b6}
\end{equation}
where $M$ is a normalization factor and
\begin{eqnarray}
W^{(\gamma)}_{\mu\nu\ \alpha\beta}&=&
{1\over 2}{\eta}_{\mu\nu}(-k_1\cdot k_2{\eta}_{\alpha\beta}+k_{1\beta}k_{2\alpha})
\label{b7}\\
&&+\ k_1\cdot k_2{\eta}_{\mu\alpha}{\eta}_{\nu\beta}
\label{b8}\\
&&-\ {\eta}_{\mu\alpha}k_{1\beta}k_{2\nu}\ -\ {\eta}_{\mu\beta}k_{1\nu}k_{2\alpha}
\label{b9}\\
&&+\ {\eta}_{\alpha\beta}k_{1\mu}k_{2\nu}
\ .
\label{b10}
\end{eqnarray}
The $X_2gg$ vertex is identical, apart from a trivial color factor $\delta^{ab}$.

We work in the $X_2$ rest frame, take the beam direction as the $z$-axis, and write the momenta
of the initial-state gluons as
\begin{equation}
k_1^{\mu}=(k_1^0,k_1^1,k_1^2,k_1^3)=(k,0,0,k)\ , \ \ \ \
k_2^{\mu}=(k_2^0,k_2^1,k_2^2,k_2^3)=(k,0,0,-k)\ .
\label{b10aa}
\end{equation}
We recall that the polarization vectors of the massless initial-state gluons are given by (\ref{a1}) and (\ref{a2}).

We denote the unit spatial vectors in the coordinate system where (\ref{b10aa}), (\ref{a1}) and (\ref{a2})
apply for the initial-state gluons by ${\hat x}$, ${\hat y}$ and ${\hat z}$.
We denote the momenta of the final-state photons by $k'_1$ and $k'_2$, and take the three-momentum vector of $k'_1$ 
to lie along the ${\hat z}'$ direction, where ${\hat x}'$, ${\hat y}'$ and ${\hat z}'$ are given by
\begin{equation}
{\hat x}'=\cos{\theta}\, {\hat x}-\sin{\theta}\, {\hat z}\ ,\ \ \
{\hat z}'=\sin{\theta}\, {\hat x}+\cos{\theta}\, {\hat z}\ ,\ \ \
{\hat y}'={\hat y}\ ,
\label{b11}
\end{equation}
i.e., ${\hat z}'$ is given by rotating ${\hat z}$ toward ${\hat x}$ by the angle ${\theta}$,
as shown in Fig. \ref{diagrams}(c).
The momenta $k'_1$ and $k'_2$, as well as the polarization vectors of the final-state photons
in the coordinate system ${\hat x}'$, ${\hat y}'$ and ${\hat z}'$, are given by
expressions identical to those given in (\ref{a1}), (\ref{a2}) and (\ref{b10aa}) for the initial-state gluons
in the coordinate system ${\hat x}$, ${\hat y}$ and ${\hat z}$.
Then, in the coordinate system ${\hat x}$, ${\hat y}$ and ${\hat z}$, we have
\begin{equation}
k_1^{\prime\mu}=k(1,\sin{\theta},0,\cos{\theta})\ , \ \ \ \
k_2^{\prime\mu}=k(1,-\sin{\theta},0,-\cos{\theta})\ ,
\label{b10aaapp2}
\end{equation}
and
\begin{equation}
\epsilon^{\prime\, +\, \mu}=
(0,-{1\over {\sqrt{2}}}\cos{\theta},-{i\over {\sqrt{2}}},{1\over {\sqrt{2}}}\sin{\theta})\ , \ \ \ \
\epsilon^{\prime\, -\, \mu}=
(0,{1\over {\sqrt{2}}}\cos{\theta},-{i\over {\sqrt{2}}},-{1\over {\sqrt{2}}}\sin{\theta})\ .
\label{b10aaapp3}
\end{equation}
The three rotated polarizations of $X_2$ are represented by the two given in (\ref{b10aaapp3}) and by
\begin{equation}
\epsilon^{\prime\, 0\, \mu}=(0,\sin{\theta} , 0, \cos{\theta})\ .
\label{b10app00}
\end{equation}
We note that the polarization vectors $\epsilon^{\prime\mu}$ are labeled by $\pm$, according to
the component of the photon spin along the $k_1^{\prime}$ direction, just as the $\epsilon^{\mu}$ are
labeled by the component of the gluon spin along the $k_1$ direction.
The amplitude of the process $gg \to X_2 \to \gamma \gamma$ is of the form
\begin{equation}
{A}(\epsilon_1^{\prime}\epsilon_2^{\prime}\, ;\, \epsilon_1\epsilon_2)
\propto \epsilon_1^{\prime a\, *}\epsilon_2^{\prime b\, *}\
W_{ab\ \rho\sigma}\
\Big( \sum_{s=-2}^{+2}\epsilon^{s\,\rho\sigma}\epsilon^{s\,\mu\nu\, *} \Big)
W_{\mu\nu\ \alpha\beta}\
\epsilon_1^{\alpha}\epsilon_2^{\beta} \ ,
\label{app1}
\end{equation}
where the vertex $W_{\mu\nu\ \alpha\beta}$ is given in (\ref{b6}).

\subsection{Calculation of Differential Cross Section}

We calculate the amplitude for $gg \to X_2 \to \gamma \gamma$ when the initial gluon
polarization state is one of $\epsilon_1^+\epsilon_2^+$, $\epsilon_1^-\epsilon_2^-$,
$\epsilon_1^+\epsilon_2^-$ and $\epsilon_1^-\epsilon_2^+$,
and the final photon polarization state is one of $\epsilon_1^{\prime\, +}\epsilon_2^{\prime\, +}$,
$\epsilon_1^{\prime\, -}\epsilon_2^{\prime\, -}$, $\epsilon_1^{\prime\, +}\epsilon_2^{\prime\, -}$
and $\epsilon_1^{\prime\, -}\epsilon_2^{\prime\, +}$ ,
via the Feynman diagram drawn in Fig. \ref{diagrams}(b).

Using the the first equation in (\ref{b2}), we see that (\ref{b7}) does not
contribute to the amplitudes. Moreover,
we see from (\ref{a1}), (\ref{a2}) and (\ref{b10aa}), that
$k_1\cdot \epsilon_2=k_2\cdot \epsilon_1=0$ and hence
(\ref{b9}) also does not contribute to the amplitudes.
Therefore, only the terms (\ref{b8}) and (\ref{b10}) in $W^{(\gamma)}_{\mu\nu\ \alpha\beta}$
may contribute to the amplitudes.

We find that the amplitude for $gg \to X_2 \to \gamma \gamma$ is non-zero only when
{\it both} of the following two conditions are satisfied:
(1) the initial gluon polarization state is one of $\epsilon_1^+\epsilon_2^+$ and $\epsilon_1^-\epsilon_2^-$,
{\it and} (2) the final photon polarization state is one of $\epsilon_1^{\prime\, +}\epsilon_2^{\prime\, +}$
and $\epsilon_1^{\prime\, -}\epsilon_2^{\prime\, -}$.
On the other hand, the amplitude is zero either when the initial gluon polarization state is one of
$\epsilon_1^+\epsilon_2^-$ and $\epsilon_1^-\epsilon_2^+$, or when
the final photon polarization state is one of $\epsilon_1^{\prime\, +}\epsilon_2^{\prime\, -}$
and $\epsilon_1^{\prime\, -}\epsilon_2^{\prime\, +}$. Thus the only possible initial and
final helicity states are $|2 2 \rangle$ and $|2 - 2 \rangle$, with no contribution from $|2 0 \rangle$.

We consider the vertex $gg \to X_2$ in the process $gg \to X_2 \to \gamma \gamma$,
which corresponds to the vertex $A$ in Fig. \ref{diagrams}(b).
When the expression (\ref{b10}), i.e., ${\eta}_{\alpha\beta}k_{1\mu}k_{2\nu}$, is attached at this vertex,
this vertex is non-zero only when the initial gluon polarization state is
$\epsilon_1^+\epsilon_2^-$ or $\epsilon_1^-\epsilon_2^+$,
since ${\eta}_{\alpha\beta}\epsilon_1^{+\, \alpha}\epsilon_2^{-\, \beta}=
{\eta}_{\alpha\beta}\epsilon_1^{-\, \alpha}\epsilon_2^{+\, \beta}=1$
and ${\eta}_{\alpha\beta}\epsilon_1^{+\, \alpha}\epsilon_2^{+\, \beta}=
{\eta}_{\alpha\beta}\epsilon_1^{-\, \alpha}\epsilon_2^{-\, \beta}=0$.
Then, using $\epsilon^{0\,\mu\nu\, *} k_{1\mu}k_{2\nu}={2\over {\sqrt{6}}}\,
(\epsilon^{0\, *}\cdot k_1)\, (\epsilon^{0\, *}\cdot k_2) = {2\over {\sqrt{6}}}\, (-k)\, (+k)
= -\, {2\over {\sqrt{6}}}\, k^2$
and $\epsilon^{(s\neq 0)\,\mu\nu\, *} k_{1\mu}k_{2\nu}=0$,
the amplitude (\ref{app1}) for this vertex becomes
\begin{eqnarray}
&&\epsilon_1^{\prime a\, *}\epsilon_2^{\prime b\, *}\
W_{ab\ \rho\sigma}\
\Big( \sum_{s=-2}^{+2}\epsilon^{s\,\rho\sigma}\epsilon^{s\,\mu\nu\, *} \Big)
\ \Big( {\eta}_{\alpha\beta}k_{1\mu}k_{2\nu} \Big)\ 
\epsilon_1^{+\, \alpha}\epsilon_2^{-\, \beta}
\nonumber\\
&=&\epsilon_1^{\prime a\, *}\epsilon_2^{\prime b\, *}\
W_{ab\ \rho\sigma}\
\Big( \sum_{s=-2}^{+2}\epsilon^{s\,\rho\sigma}\epsilon^{s\,\mu\nu\, *} \Big)
\ \Big( {\eta}_{\alpha\beta}k_{1\mu}k_{2\nu} \Big)\ 
\epsilon_1^{-\, \alpha}\epsilon_2^{+\, \beta}
\nonumber\\
&=&\epsilon_1^{\prime a\, *}\epsilon_2^{\prime b\, *}\
W_{ab\ \rho\sigma}\
\epsilon^{0\,\rho\sigma}\ \Big( -\, {2\over {\sqrt{6}}}\, k^2\Big)
\ ,
\label{app1a}
\end{eqnarray}
and
\begin{eqnarray}
&&\epsilon_1^{\prime a\, *}\epsilon_2^{\prime b\, *}\
W_{ab\ \rho\sigma}\
\Big( \sum_{s=-2}^{+2}\epsilon^{s\,\rho\sigma}\epsilon^{s\,\mu\nu\, *} \Big)
\ \Big( {\eta}_{\alpha\beta}k_{1\mu}k_{2\nu} \Big)\ 
\epsilon_1^{+\, \alpha}\epsilon_2^{+\, \beta}
\nonumber\\
&=&\epsilon_1^{\prime a\, *}\epsilon_2^{\prime b\, *}\
W_{ab\ \rho\sigma}\
\Big( \sum_{s=-2}^{+2}\epsilon^{s\,\rho\sigma}\epsilon^{s\,\mu\nu\, *} \Big)
\ \Big( {\eta}_{\alpha\beta}k_{1\mu}k_{2\nu} \Big)\ 
\epsilon_1^{-\, \alpha}\epsilon_2^{-\, \beta}
\nonumber\\
&=&0
\ .
\label{app1b}
\end{eqnarray}
When the expression (\ref{b8}), i.e., $k_1\cdot k_2{\eta}_{\mu\alpha}{\eta}_{\nu\beta}$,
is attached at this vertex,
using $\epsilon^{+\, *}\cdot \epsilon_1^{+}=\epsilon^{-\, *}\cdot \epsilon_1^{-}
=\epsilon^{+\, *}\cdot \epsilon_2^{+}=\epsilon^{-\, *}\cdot \epsilon_2^{-}=-\, 1$
and $k_1\cdot k_2=2k^2$, we find
\begin{eqnarray}
&&\epsilon_1^{\prime a\, *}\epsilon_2^{\prime b\, *}\
W_{ab\ \rho\sigma}\
\Big( \sum_{s=-2}^{+2}\epsilon^{s\,\rho\sigma}\epsilon^{s\,\mu\nu\, *} \Big)
\ \Big( k_1\cdot k_2{\eta}_{\mu\alpha}{\eta}_{\nu\beta} \Big)\ 
\epsilon_1^{+\, \alpha}\epsilon_2^{-\, \beta}
\nonumber\\
&=&\epsilon_1^{\prime a\, *}\epsilon_2^{\prime b\, *}\
W_{ab\ \rho\sigma}\
\Big( \sum_{s=-2}^{+2}\epsilon^{s\,\rho\sigma}\epsilon^{s\,\mu\nu\, *} \Big)
\ \Big( k_1\cdot k_2{\eta}_{\mu\alpha}{\eta}_{\nu\beta} \Big)\ 
\epsilon_1^{-\, \alpha}\epsilon_2^{+\, \beta}
\nonumber\\
&=&\epsilon_1^{\prime a\, *}\epsilon_2^{\prime b\, *}\
W_{ab\ \rho\sigma}\
\epsilon^{0\,\rho\sigma}\ \Big( +\, {1\over {\sqrt{6}}}\, 2k^2\Big)
\ ,
\label{app1c}
\end{eqnarray}
and
\begin{eqnarray}
\epsilon_1^{\prime a\, *}\epsilon_2^{\prime b\, *}\
W_{ab\ \rho\sigma}\
\Big( \sum_{s=-2}^{+2}\epsilon^{s\,\rho\sigma}\epsilon^{s\,\mu\nu\, *} \Big)
\ \Big( k_1\cdot k_2{\eta}_{\mu\alpha}{\eta}_{\nu\beta} \Big)\ 
\epsilon_1^{+\, \alpha}\epsilon_2^{+\, \beta}
&=&
\epsilon_1^{\prime a\, *}\epsilon_2^{\prime b\, *}\
W_{ab\ \rho\sigma}\
\epsilon^{+2\,\rho\sigma}\ \Big( 2k^2\Big)
\nonumber\\
\epsilon_1^{\prime a\, *}\epsilon_2^{\prime b\, *}\
W_{ab\ \rho\sigma}\
\Big( \sum_{s=-2}^{+2}\epsilon^{s\,\rho\sigma}\epsilon^{s\,\mu\nu\, *} \Big)
\ \Big( k_1\cdot k_2{\eta}_{\mu\alpha}{\eta}_{\nu\beta} \Big)\ 
\epsilon_1^{-\, \alpha}\epsilon_2^{-\, \beta}
&=&\epsilon_1^{\prime a\, *}\epsilon_2^{\prime b\, *}\
W_{ab\ \rho\sigma}\
\epsilon^{-2\,\rho\sigma}\ \Big( 2k^2\Big)
\ .\qquad
\label{app1d}
\end{eqnarray}
Combining (\ref{app1a}) to (\ref{app1d}), we have
\begin{eqnarray}
&&\epsilon_1^{\prime a\, *}\epsilon_2^{\prime b\, *}\
W_{ab\ \rho\sigma}\
\Big( \sum_{s=-2}^{+2}\epsilon^{s\,\rho\sigma}\epsilon^{s\,\mu\nu\, *} \Big)
\ \Big( k_1\cdot k_2{\eta}_{\mu\alpha}{\eta}_{\nu\beta} \ +\
{\eta}_{\alpha\beta}k_{1\mu}k_{2\nu} \Big)\ 
\epsilon_1^{+\, \alpha}\epsilon_2^{-\, \beta}
\nonumber\\
&=&\epsilon_1^{\prime a\, *}\epsilon_2^{\prime b\, *}\
W_{ab\ \rho\sigma}\
\Big( \sum_{s=-2}^{+2}\epsilon^{s\,\rho\sigma}\epsilon^{s\,\mu\nu\, *} \Big)
\ \Big( k_1\cdot k_2{\eta}_{\mu\alpha}{\eta}_{\nu\beta} \ +\
{\eta}_{\alpha\beta}k_{1\mu}k_{2\nu} \Big)\ 
\epsilon_1^{-\, \alpha}\epsilon_2^{+\, \beta}
\nonumber\\
&=&0
\ ,
\label{app1e}
\end{eqnarray}
and
\begin{eqnarray}
&&\epsilon_1^{\prime a\, *}\epsilon_2^{\prime b\, *}\
W_{ab\ \rho\sigma}\
\Big( \sum_{s=-2}^{+2}\epsilon^{s\,\rho\sigma}\epsilon^{s\,\mu\nu\, *} \Big)
\ \Big( k_1\cdot k_2{\eta}_{\mu\alpha}{\eta}_{\nu\beta} \ +\
{\eta}_{\alpha\beta}k_{1\mu}k_{2\nu} \Big)\ 
\epsilon_1^{+\, \alpha}\epsilon_2^{+\, \beta}
\nonumber\\
&&\qquad\qquad =\
\epsilon_1^{\prime a\, *}\epsilon_2^{\prime b\, *}\
W_{ab\ \rho\sigma}\
\epsilon^{+2\,\rho\sigma}\ \Big( 2k^2\Big) \ ,
\nonumber\\
&&\epsilon_1^{\prime a\, *}\epsilon_2^{\prime b\, *}\
W_{ab\ \rho\sigma}\
\Big( \sum_{s=-2}^{+2}\epsilon^{s\,\rho\sigma}\epsilon^{s\,\mu\nu\, *} \Big)
\ \Big( k_1\cdot k_2{\eta}_{\mu\alpha}{\eta}_{\nu\beta} \ +\
{\eta}_{\alpha\beta}k_{1\mu}k_{2\nu} \Big)\ 
\epsilon_1^{-\, \alpha}\epsilon_2^{-\, \beta}
\nonumber\\
&&\qquad\qquad =\
\epsilon_1^{\prime a\, *}\epsilon_2^{\prime b\, *}\
W_{ab\ \rho\sigma}\
\epsilon^{-2\,\rho\sigma}\ \Big( 2k^2\Big)
\ .\qquad
\label{app1f}
\end{eqnarray}
Equivalent results are obtained when the roles of the initial gluon polarization states
and the final photon polarization states are exchanged in (\ref{app1e}) and (\ref{app1f}).

This analysis justifies the statement made at the beginning of this subsection,
namely that the amplitude for $gg \to X_2 \to \gamma \gamma$ is non-zero only when
{\it both} of the following two conditions are satisfied:
(1) the initial gluon polarization state is one of $\epsilon_1^+\epsilon_2^+$ and $\epsilon_1^-\epsilon_2^-$,
{\it and} (2) the final photon polarization state is one of $\epsilon_1^{\prime\, +}\epsilon_2^{\prime\, +}$
and $\epsilon_1^{\prime\, -}\epsilon_2^{\prime\, -}$.
That is, the amplitude is zero {\it either} when the initial gluon polarization state is one of
$\epsilon_1^+\epsilon_2^-$ and $\epsilon_1^-\epsilon_2^+$, {\it or} when
the final photon polarization state is one of $\epsilon_1^{\prime\, +}\epsilon_2^{\prime\, -}$
and $\epsilon_1^{\prime\, -}\epsilon_2^{\prime\, +}$.

We also see in (\ref{app1f}) that in the non-zero amplitude found when
both (\ref{b8}) and (\ref{b10}) are attached at both vertices in the Feynman diagram
is the same as that obtained when only (\ref{b8}) is attached at both vertices
in the Feynman diagram.

\begin{figure}
\centering
\begin{minipage}[t]{10.0cm}
\centering
\includegraphics[width=\textwidth]{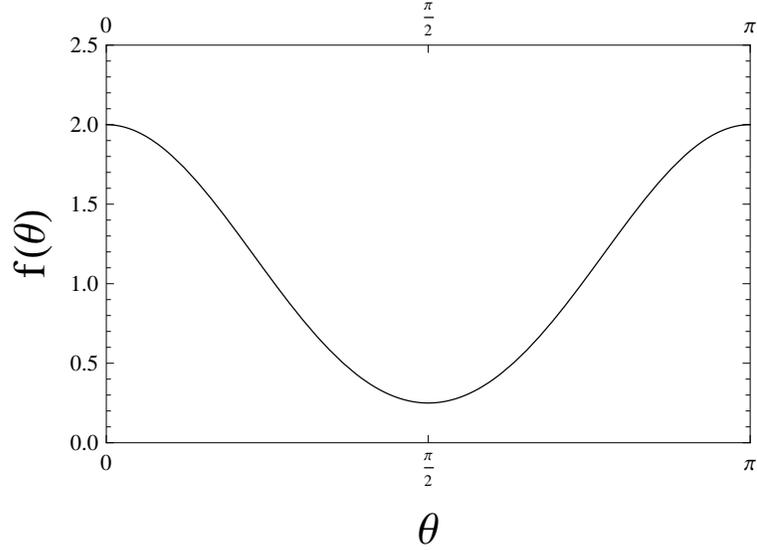}
\end{minipage}\hspace{1.0cm}
\parbox{0.95\textwidth}{\caption{
\it The $\gamma \gamma$ angular distribution of $d \sigma / d \Omega$ given in (\protect\ref{g10}).
\label{comparison5a}}}
\end{figure}

When the sum of (\ref{b8}) and (\ref{b10}), i.e.,
$(k_1\cdot k_2{\eta}_{\mu\alpha}{\eta}_{\nu\beta}\ +\ {\eta}_{\alpha\beta}k_{1\mu}k_{2\nu})$,
is attached at both vertices in the Feynman diagram for
$gg \to X_2 \to \gamma \gamma$ shown in Fig. \ref{diagrams}(b),
using (\ref{a1}), (\ref{a2}), (\ref{b1}) and (\ref{b10aaapp3})
in (\ref{app1f}),
we find that
the amplitudes (\ref{app1}) are given by $(4k^2 / M)^2$ times the following angular expressions:
\begin{eqnarray}
A(+'\, +'\, ; +\, +)&=&A(-'\, -'\, ; -\, -)\ =\ {1\over 4}(1+\cos{\theta})^2
\label{app11a}\\
A(-'\, -'\, ; +\, +)&=&A(+'\, +'\, ; -\, -)\ =\ {1\over 4}(1-\cos{\theta})^2
\label{app12a}\\
A(+'\, -'\, ; +\, +)&=&A(-'\, +'\, ; +\, +)\ =\ A(+'\, -'\, ; -\, -)\
=\ \ A(-'\, +'\, ; -\, -)\ =\ 0
\label{app13a}\\
A(+'\, +'\, ; +\, -)&=&A(-'\, -'\, ; +\, -)\ =\ A(+'\, +'\, ; -\, +)\
=\ \ A(-'\, -'\, ; -\, +)\ =\ 0
\label{app14a}\\
A(+'\, -'\, ; +\, -)&=&A(-'\, +'\, ; +\, -)\ =\ A(+'\, -'\, ; -\, +)\
=\ \ A(-'\, +'\, ; -\, +)\ =\ 0
\ .\qquad
\label{app15a}
\end{eqnarray}
The contributions
of the two possible final polarization states
$\epsilon_1^{\prime\, +}\epsilon_2^{\prime\, +}$ and $\epsilon_1^{\prime\, -}\epsilon_2^{\prime\, -}$
to the total $\gamma \gamma$ cross section ${d\sigma/d\Omega}$
are identical, and we have (as derived earlier in~\cite{10}):
\begin{equation}
{d\sigma \over d\Omega}\ \propto\ {1\over 4} + {3\over 2}{\rm cos}^2{\theta}
+ {1\over 4}{\rm cos}^4{\theta}
\ ,
\label{g10}
\end{equation}
which is plotted in Fig. \ref{comparison5a}.

We see in Fig. \ref{comparison5a} that the total $\gamma \gamma$ angular distribution in the $X_2$
centre-of-mass frame differs substantially from the isotropic angular distribution expected for
the decay of a spin-zero particle such as the Higgs boson. 
In particular, the $\gamma \gamma$ final state is suppressed at large angles $\theta$
relative to the beams. This suggests that a careful study
of the $\gamma \gamma$ angular distribution might offer some discrimination between the
spin-two and spin-zero hypotheses. Any conclusion on this possibility would
require a realistic simulation of the $\gamma \gamma$ signal in an LHC detector.
However, we estimate that the centre-of-mass system of a photon pair can be
reconstructed quite accurately, the dominant uncertainties probably being due to errors in the photon
energy measurements, which are at the 1\% level in both ATLAS and CMS. The preliminary
results of simulation studies using {\tt Pythia} and {\tt Delphes}~\cite{EFHSY} support the expectation that
the $\gamma \gamma$ angular distribution is little affected by detector effects.

\section{The Process $gg \to X_2 \to W^- W^+\to \ell^- \ell^+ {\bar \nu} \nu$}

\subsection{Lepton Angular Distributions in $W$ Decays}

\subsubsection{$W^- \to \ell^- {\bar \nu}$}

\begin{figure}
\centering
\vspace*{-0.5cm}
\begin{minipage}[t]{8.5cm}
\centering
\includegraphics[width=\textwidth]{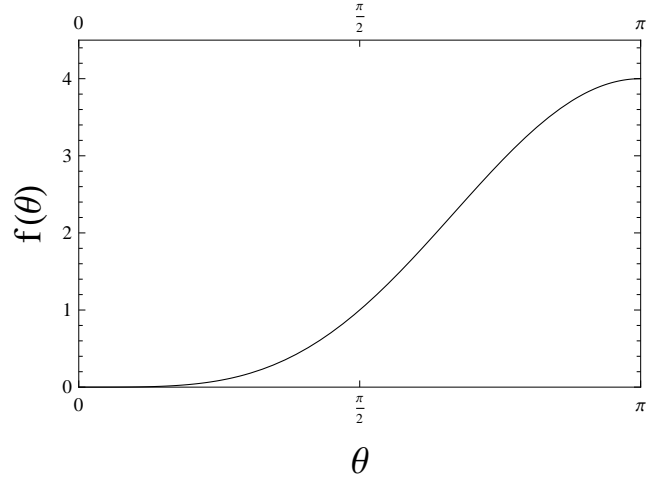}
(a)
\end{minipage}\hspace{1.0cm}
\begin{minipage}[t]{8.5cm}
\centering
\includegraphics[width=\textwidth]{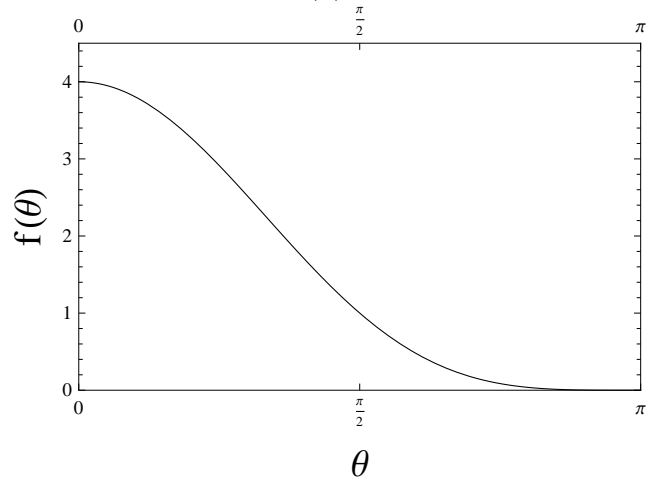}
(b)
\end{minipage}\hspace{1.0cm}
\begin{minipage}[t]{8.5cm}
\centering
\includegraphics[width=\textwidth]{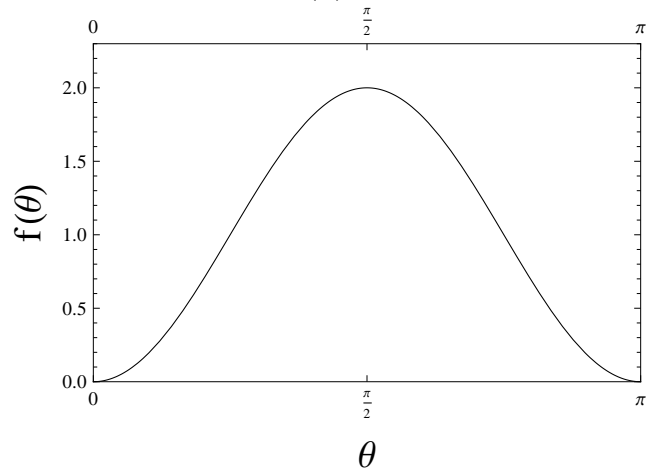}
(c)
\end{minipage}\hspace{0.0cm}
\parbox{0.95\textwidth}{\caption{
\it The decay angular distribution functions $f(\theta)$ in $W^- \to \ell^- {\bar \nu}$ decays
from the $W^-$ polarization states given by (a) $\epsilon^{+}$,
(b) $\epsilon^{-}$, and (c) $\epsilon^{0}$.
\label{fig:comparison2}}}
\end{figure}

As preparation for this Section,
we first consider the decay $W^- \to \ell^- {\bar \nu}$.
We consider a $W^-$ at rest and denote the momenta of the final-state particles by 
\begin{eqnarray}
p_{\ell^-}^\mu &=& (p,p\sin{\theta_1}\cos{\phi_1},p\sin{\theta_1}\sin{\phi_1},p\cos{\theta_1})\ ,
\label{aaa1wa1}\\
p_{{\bar \nu}}^\mu &=& (p,-p\sin{\theta_1}\cos{\phi_1},-p\sin{\theta_1}\sin{\phi_1},-p\cos{\theta_1})\ ,
\label{aaa1wa2}
\end{eqnarray}
where $p=|{\vec p}|$ and the $\ell^-$ mass is ignored.
The polarization vectors $\epsilon$ of $W^-$ with the $z$-axis as the quantization axis
are given by (\ref{a1}, \ref{a2}) and $\epsilon^{0\mu} =(0, 0, 0, 1)$.
We calculate
${\cal M}_1={\bar u}(p_{\ell^-})\gamma^\mu \epsilon_{1\mu} (1-\gamma_5)v(p_{{\bar \nu}})$
and find the following results for ${\cal M}_1/(2{\sqrt{2}}\, p)$:
\begin{eqnarray}
&&{\rm for}\ \epsilon_1^{+}\ ,\ \ \ \ \ \
(1-\cos{\theta_1})\ e^{+i\phi_1}
\label{a4w}\\
&&{\rm for}\ \epsilon_1^{-}\ ,\ \ \ \ \ \
(1+\cos{\theta_1})\ e^{-i\phi_1}
\label{a5w}\\
&&{\rm for}\ \epsilon_1^{0}\ ,\ \ \ \ \ \
-\, {\sqrt{2}}\ \sin{\theta_1} \ .
\label{a6w}
\end{eqnarray}
The differential cross section ${d\sigma / d\Omega}$ is proportional to $|{\cal M}_1|^2$
and the functions $f(\theta) = |{\cal M}_1/(2{\sqrt{2}}\, p)|^2$  for the three polarization
states are plotted in Fig. \ref{fig:comparison2}.

\subsubsection{$W^+ \to \ell^+ {\nu}$}

In the case of $W^+$ at rest, we denote the momenta of the final-state particles by
\begin{eqnarray}
p_{\ell^+}^\mu &=& (p,p\sin{\theta_2}\cos{\phi_2},p\sin{\theta_2}\sin{\phi_2},p\cos{\theta_2})\ ,
\label{aaa1wa1lp}\\
p_{{\nu}}^\mu &=& (p,-p\sin{\theta_2}\cos{\phi_2},-p\sin{\theta_2}\sin{\phi_2},-p\cos{\theta_2})\ .
\label{aaa1wa2lp}
\end{eqnarray}
We calculate
${\cal M}_2={\bar u}(p_{\nu})\gamma^\mu \epsilon_{2\mu} (1-\gamma_5)v(p_{\ell^+})$
and find the following results for ${\cal M}_2/(2{\sqrt{2}}\, p)$:
\begin{eqnarray}
&&{\rm for}\ \epsilon_2^{+}\ ,\ \ \ \ \ \
-\ (1+\cos{\theta_2})\ e^{+i\phi_2}
\label{a4wlp}\\
&&{\rm for}\ \epsilon_2^{-}\ ,\ \ \ \ \ \
-\ (1-\cos{\theta_2})\ e^{-i\phi_2}
\label{a5wlp}\\
&&{\rm for}\ \epsilon_2^{0}\ ,\ \ \ \ \ \ \ \
-\, {\sqrt{2}}\ \sin{\theta_2} \ .
\label{a6wlp}
\end{eqnarray}

\subsection{Angular Distributions in $gg \to X_2 (X_0) \to W^- W^+ \to \ell^- \ell^+ {\bar \nu} \nu$}

For simplicity, we assume that the $W^-$ and $W^+$ are at rest, corresponding to the case
that $m_X = 2 m_W$. In practice, we are interested in the decay of the possible particle with
mass $\sim 125$~GeV reported by ATLAS and CMS, which would decay into one on-shell $W$
and one off-shell $W^* \to \ell \nu$. The structure of the $W \ell \nu$ decay matrix element would 
be dominated by the $W^*$ pole, favouring $\ell \nu$ invariant masses close to $m_X - m_W$
and hence small momenta for the $W$ and $W^*$ in the centre-of-mass frame of the decaying $X$
particle. The crude approximation of neglecting these momenta may serve to indicate whether
in principle there could be significant differences between the decay angular distributions in
$X_2$ and $X_0$ decay that could be investigated in more detailed simulations.

With this assumption, we denote the polarization vector and momentum of the $W^-$ ($W^+$) by
$\epsilon_1$ and $k_1$ ($\epsilon_2$ and $k_2$), respectively.
The polarization vectors $\epsilon_1$ and $\epsilon_2$ with the $z$-axis as the quantization axis
are given by (\ref{a1}, \ref{a2}) and $\epsilon^{0\mu} =(0, 0, 0, 1)$ as before,
and the momenta $k_1$ and $k_2$ are given by
\begin{equation}
k_1^{\mu}=k_2^{\mu}=(k_1^0,k_1^1,k_1^2,k_1^3)=(m_W,0,0,0)=({m\over 2},0,0,0)\ .
\label{a3pa}
\end{equation}
For the Feynman rule of the three-point vertex $X_2W^-W^+$,
we use the following vertex which is given in \cite{Han}:
\begin{equation}
-\, {i\over {\bar M}_P}\, \Big( W^{(W)}_{\mu\nu\ \alpha\beta}\ +\
W^{(W)}_{\nu\mu\ \alpha\beta}\Big)\ ,
\label{b6WW}
\end{equation}
where
\begin{eqnarray}
W^{(W)}_{\mu\nu\ \alpha\beta}&=&
{1\over 2}{\eta}_{\mu\nu}(-(m_W^2+k_1\cdot k_2){\eta}_{\alpha\beta}+k_{1\beta}k_{2\alpha})
\label{b7a}\\
&&+\ (m_W^2+k_1\cdot k_2){\eta}_{\mu\alpha}{\eta}_{\nu\beta}
\label{b8a}\\
&&-\ {\eta}_{\mu\alpha}k_{1\beta}k_{2\nu}\ -\ {\eta}_{\mu\beta}k_{1\nu}k_{2\alpha}
\label{b9a}\\
&&+\ {\eta}_{\alpha\beta}k_{1\mu}k_{2\nu}
\ .
\label{b10a}
\end{eqnarray}
Since $m_W^2+k_1\cdot k_2 = ({m\over 2})^2+({m\over 2})^2={m^2\over 2}$,
we may write $W^{(W)}_{\mu\nu\ \alpha\beta}$ above as
\begin{eqnarray}
W^{(W)}_{\mu\nu\ \alpha\beta}&=&
{1\over 2}{\eta}_{\mu\nu}(-{m^2\over 2}{\eta}_{\alpha\beta}+k_{1\beta}k_{2\alpha})
\label{b7b}\\
&&+\ {m^2\over 2}{\eta}_{\mu\alpha}{\eta}_{\nu\beta}
\label{b8b}\\
&&-\ {\eta}_{\mu\alpha}k_{1\beta}k_{2\nu}\ -\ {\eta}_{\mu\beta}k_{1\nu}k_{2\alpha}
\label{b9b}\\
&&+\ {\eta}_{\alpha\beta}k_{1\mu}k_{2\nu}
\ .
\label{b10b}
\end{eqnarray}
When we work with the simplified kinematical case (\ref{a3pa}), only the second line (\ref{b8b}) of the above
expression for $W^{(W)}_{\mu\nu\ \alpha\beta}$ contributes in the present calculation.

We calculate the angular distributions of the $\ell^-$ and $\ell^+$ for each of the possible initial gluon
polarization states. We work in the $X_2$ rest frame, take the beam direction as the $z$-axis and write the gluon momenta as
\begin{equation}
k_1^{\mu}=(k_1^0,k_1^1,k_1^2,k_1^3)=(k,0,0,k)\ , \ \ \ \
k_2^{\mu}=(k_1^0,k_1^1,k_1^2,k_1^3)=(k,0,0,-k)\ .
\label{b10aaa}
\end{equation}
The polarization vectors of the initial gluons are given by (\ref{a1}) and (\ref{a2}), and
we denote the polarization of the gluon which has the momentum
$k_1$ ($k_2$) in (\ref{b10aaa}) by $\epsilon^g_1$ ($\epsilon^g_2$).

\subsection{Angular Correlations in $gg \to X_2 \to W^- W^+ \to \ell^- \ell^+ {\bar \nu} \nu$}

We consider the decays $X_2 \to W^- W^+ \to \ell^- \ell^+ {\bar \nu} \nu$ following $X_2$
production by $gg$ collisions with polarizations $\epsilon^g_1= \epsilon^\pm, \epsilon^g_2=\epsilon^\pm$.
Collisions with $\epsilon^g_1= \epsilon^+, \epsilon^g_2=\epsilon^+$ produce the $X_2$ in a
$|JJ^z\rangle=|2 +2 \rangle$ state, whereas collisions with $\epsilon^g_1= \epsilon^-, \epsilon^g_2=\epsilon^-$ 
produce the $X_2$ in a $|JJ^z\rangle=|2 -2 \rangle$ state. As we saw in Section~3.2, collisions with
$\epsilon^g_1= \epsilon^+, \epsilon^g_2=\epsilon^-$ and
$\epsilon^g_1= \epsilon^-, \epsilon^g_2=\epsilon^+$ have vanishing amplitudes for producing
the polarization state $|JJ^z\rangle=|20\rangle$ of the $X_2$.

\subsubsection{$|JJ^z\rangle=|2+2\rangle$}

\begin{figure}
\centering
\psfrag{xperp}[cc][cc]{$x_\perp$}
\vspace*{-0.5cm}
\begin{minipage}[t]{7.3cm}
\centering
\includegraphics[width=\textwidth]{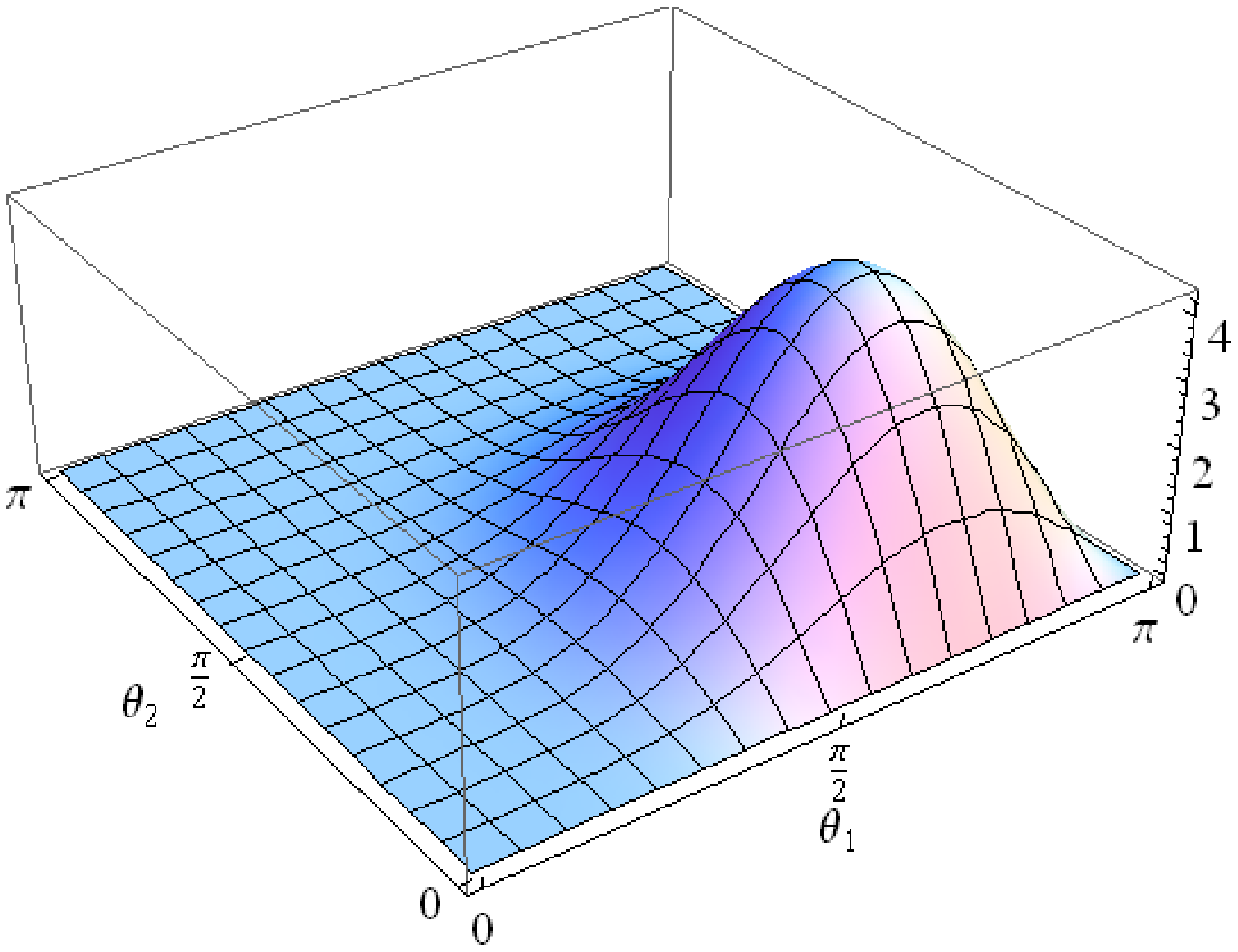}
(a1)
\end{minipage}\hspace{1.0cm}
\begin{minipage}[t]{6.7cm}
\centering
\includegraphics[width=\textwidth]{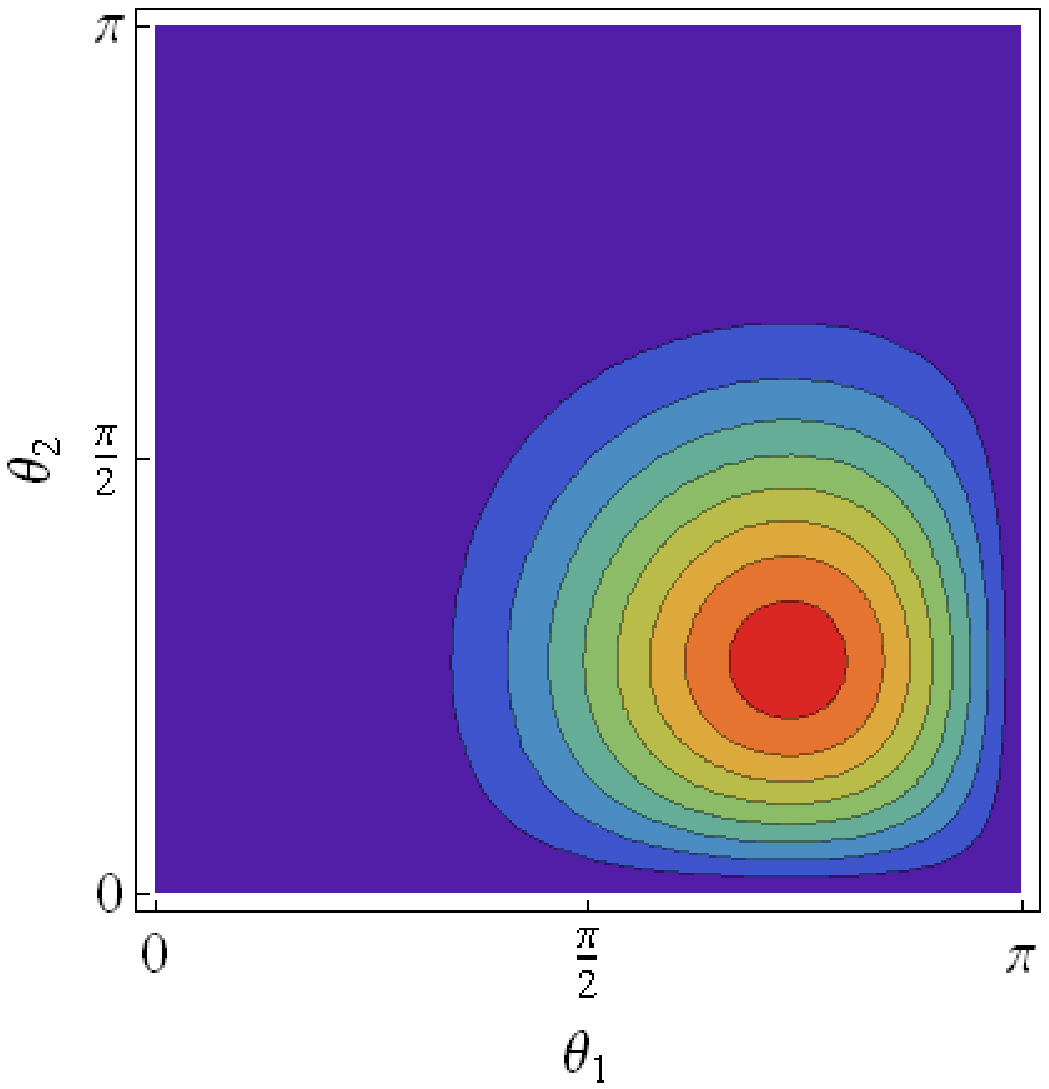}
(a2)
\end{minipage}\hspace{0.0cm}
\begin{minipage}[t]{7.3cm}
\centering
\includegraphics[width=\textwidth]{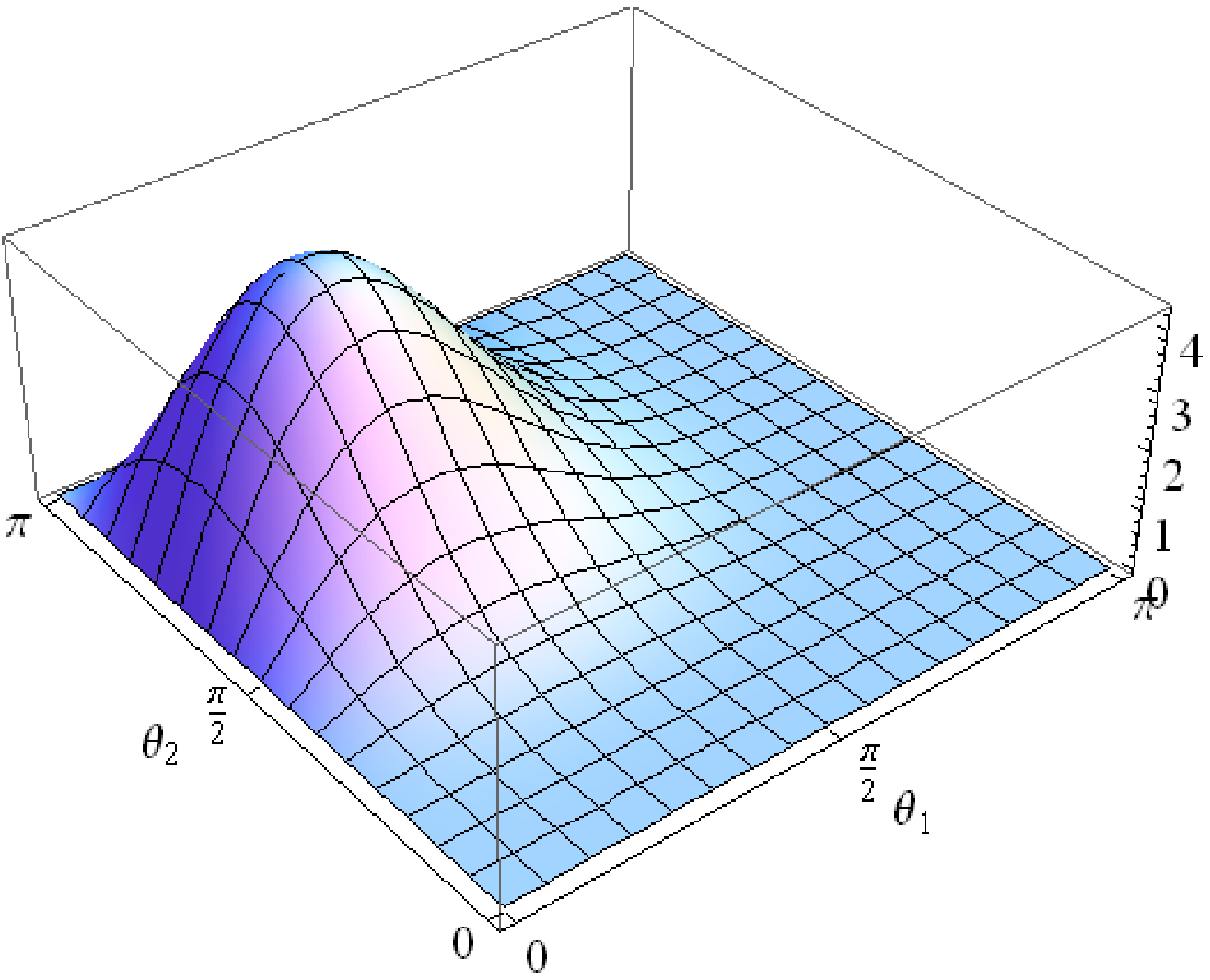}
(b1)
\end{minipage}\hspace{1.0cm}
\begin{minipage}[t]{6.7cm}
\centering
\includegraphics[width=\textwidth]{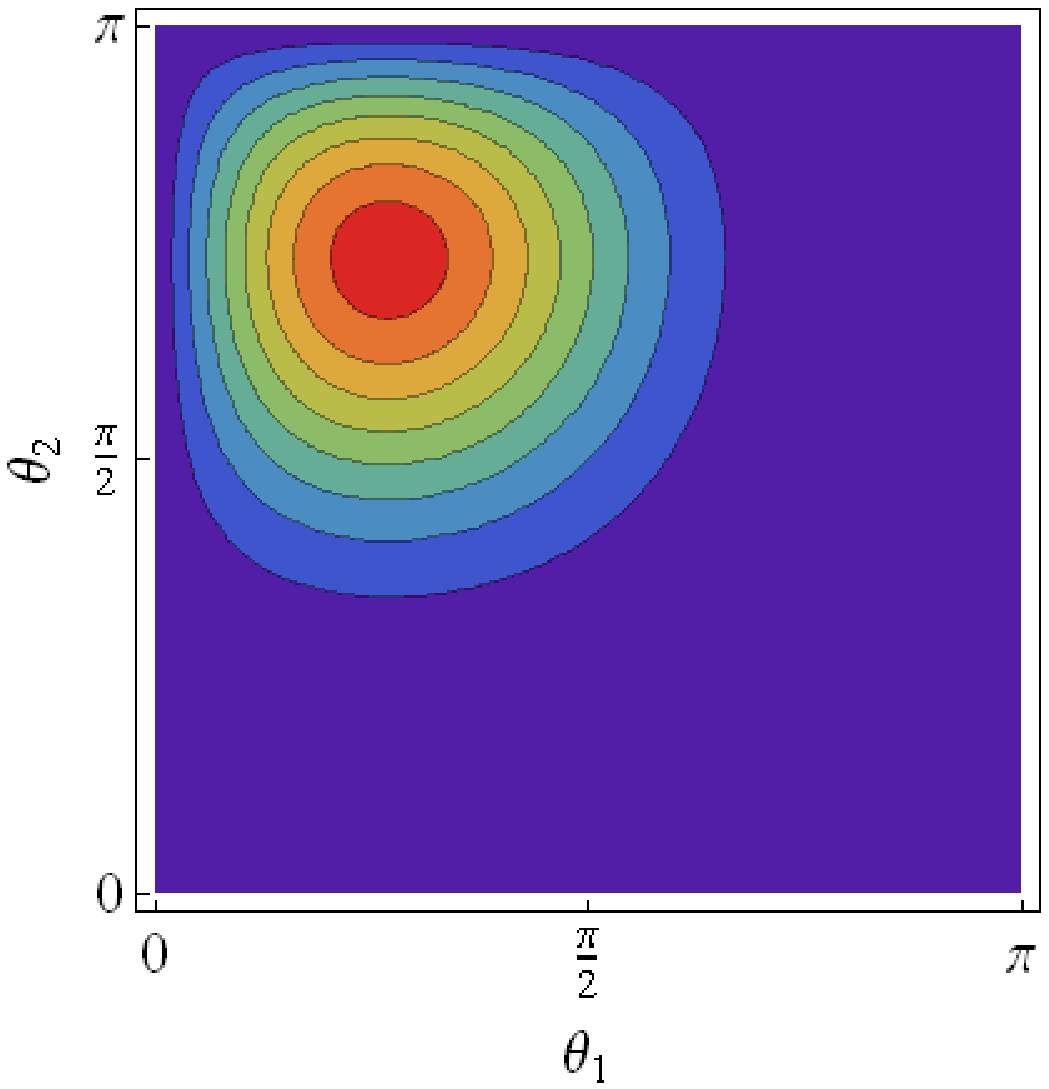}
(b2)
\end{minipage}\hspace{0.0cm}
\begin{minipage}[t]{7.3cm}
\centering
\includegraphics[width=\textwidth]{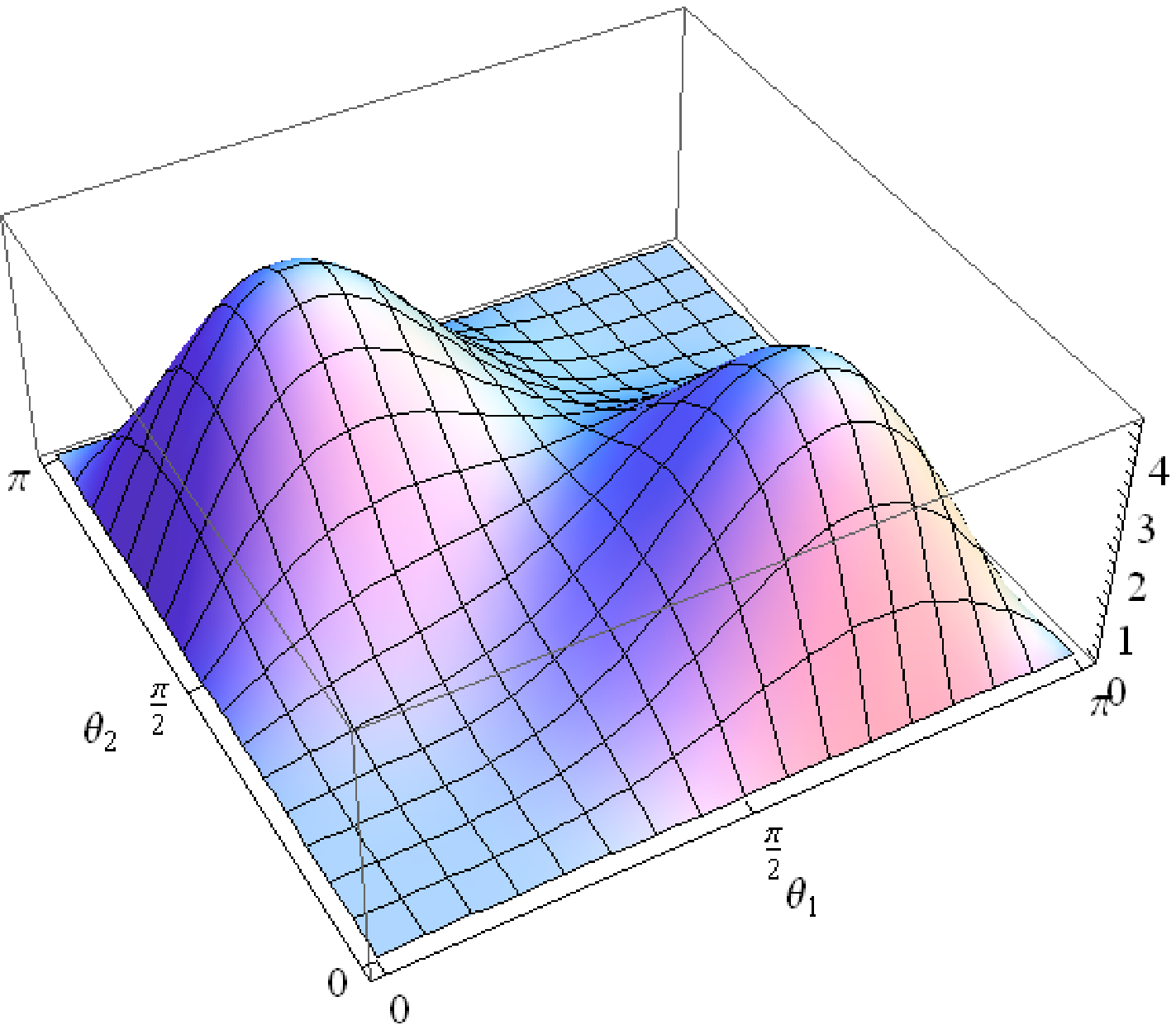}
(c1)
\end{minipage}\hspace{1.0cm}
\begin{minipage}[t]{6.7cm}
\centering
\includegraphics[width=\textwidth]{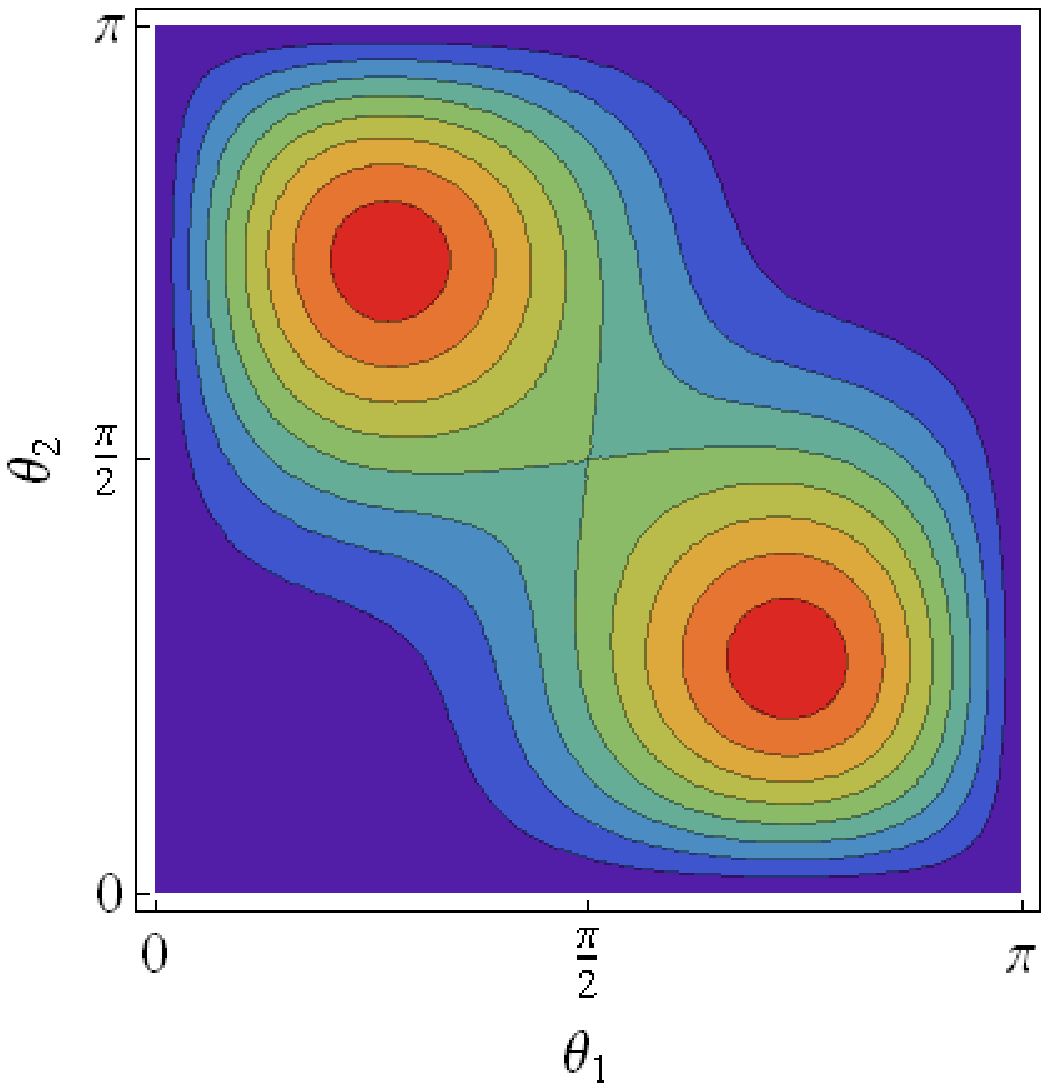}
(c2)
\end{minipage}\hspace{0.0cm}
\parbox{0.95\textwidth}{\caption{
\it The angular distributions given by (a) (\protect\ref{wwplusplusmmsquare}) $\times \sin{\theta_1}\, \sin{\theta_2}$
for decays of the $|JJ^z\rangle=|2+2\rangle$ state of $X_2 \to W^- W^+ \to \ell^- \ell^+ {\bar \nu} \nu$,
(b) (\protect\ref{wwminusminusmmsquare}) $\times \sin{\theta_1}\, \sin{\theta_2}$
for decays of the $|JJ^z\rangle=|2-2\rangle$ state of $X_2 \to W^- W^+ \to \ell^- \ell^+ {\bar \nu} \nu$,
and
(c) the sum of (a) and (b).
\label{fig:WWppmm}}}
\end{figure}

When the polarizations of the initial gluons are given by
$\epsilon^g_1= \epsilon^g_2=\epsilon^+$ and the initial
two-gluon polarization state is $|2+2\rangle$,
the polarizations of $W^-$ and $W^+$ are also given by
$\epsilon_1=\epsilon^{+}$ and $\epsilon_1=\epsilon^{+}$.
The amplitude
\begin{equation}
{\cal M}={\cal M}_1{\cal M}_2
=\Big( {\bar u}(p_{\ell^-})\gamma^\mu \epsilon_{1\mu} (1-\gamma_5)v(p_{{\bar \nu}})\Big)\,
\Big( {\bar u}(p_{\nu})\gamma^\mu \epsilon_{2\mu} (1-\gamma_5)v(p_{\ell^+})\Big)\ ,
\label{aab1w1lmlp}
\end{equation}
is then, from (\ref{a4w}) and (\ref{a4wlp}), given by:
\begin{equation}
\frac{\cal M}{(2{\sqrt{2}}\, p)^2} \, = \, -\ (1-\cos{\theta_1})\ (1+\cos{\theta_2})\ e^{i(\phi_1 +\phi_2)} \ ,
\label{wwplusplusmm}
\end{equation}
whose absolute square is independent of the azimuthal angles $\phi_{1,2}$ and proportional to:
\begin{equation}
(1-\cos{\theta_1})^2\ (1+\cos{\theta_2})^2 \ .
\label{wwplusplusmmsquare}
\end{equation}
In Fig.~\ref{fig:WWppmm}(a) we plot the quantity
(\ref{wwplusplusmmsquare}) multiplied by $\sin{\theta_1}\, \sin{\theta_2}$,
to which ${d^2\sigma / d\theta_1d\theta_2}$ is proportional.

\subsubsection{$|JJ^z\rangle=|2-2\rangle$}

Similarly, when the polarizations of the initial gluons are given by
$\epsilon^g_1= \epsilon^g_2=\epsilon^-$ and the initial
two-gluon polarization state is $|2-2\rangle$,
the polarizations of $W^-$ and $W^+$ are also given by
$\epsilon_1=\epsilon^{-}$ and $\epsilon_1=\epsilon^{-}$, and
from (\ref{a5w}) and (\ref{a5wlp})
the final-state lepton-antilepton angular distribution is again independent of the azimuthal angles $\phi_{1,2}$ and proportional to
\begin{equation}
(1+\cos{\theta_1})^2\ (1-\cos{\theta_2})^2 \ .
\label{wwminusminusmmsquare}
\end{equation}
In Fig.~\ref{fig:WWppmm}(b) we plot the quantity
(\ref{wwminusminusmmsquare}) times $\sin{\theta_1}\, \sin{\theta_2}$,
to which ${d^2\sigma / d\theta_1d\theta_2}$ is proportional in this case.
The sum of Figs. \ref{fig:WWppmm}(a) and (b) is plotted in Fig. \ref{fig:WWppmm}(c).


\subsection{Angular Correlations in $gg \to X_0 \to W^- W^+ \to \ell^- \ell^+ {\bar \nu} \nu$}

\begin{figure}
\centering
\psfrag{xperp}[cc][cc]{$x_\perp$}
\vspace*{-0.5cm}
\begin{minipage}[t]{7.3cm}
\centering
\includegraphics[width=\textwidth]{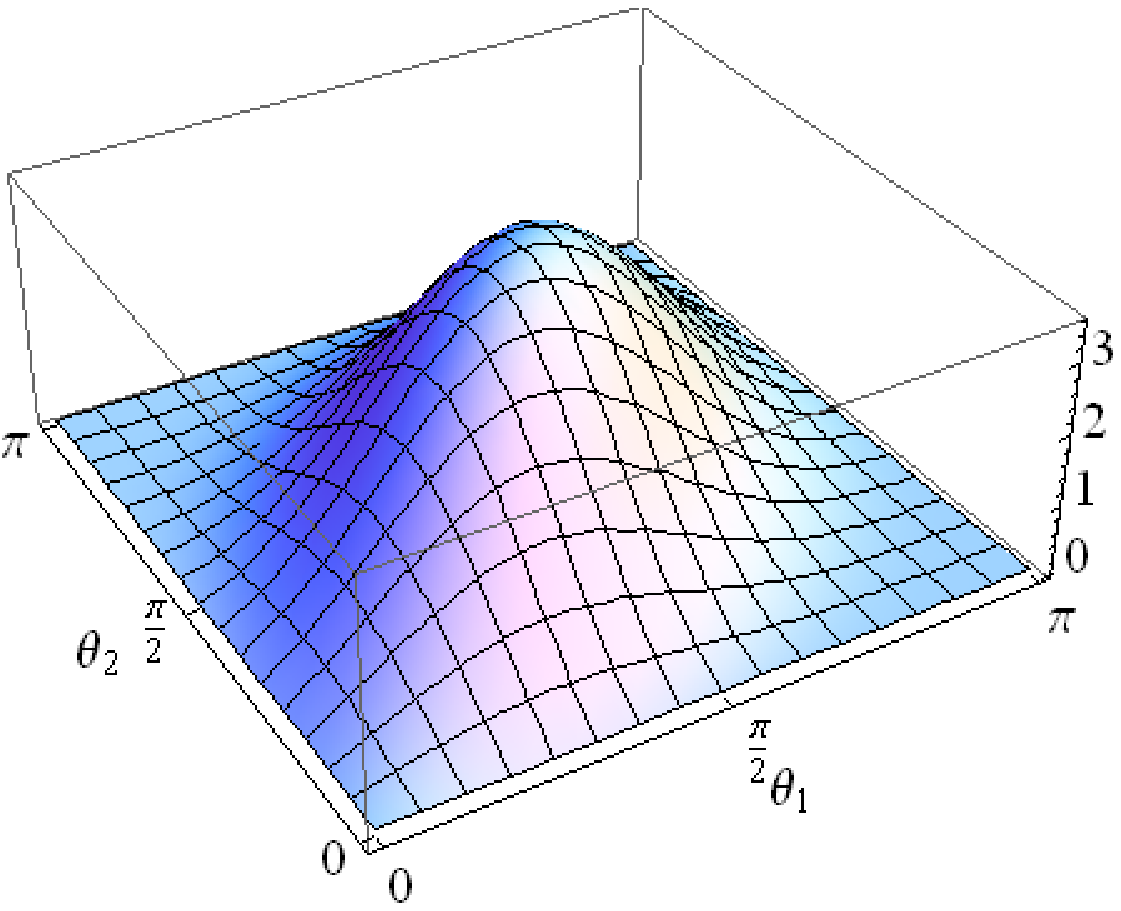}
(a1)
\end{minipage}\hspace{1.0cm}
\begin{minipage}[t]{6.7cm}
\centering
\includegraphics[width=\textwidth]{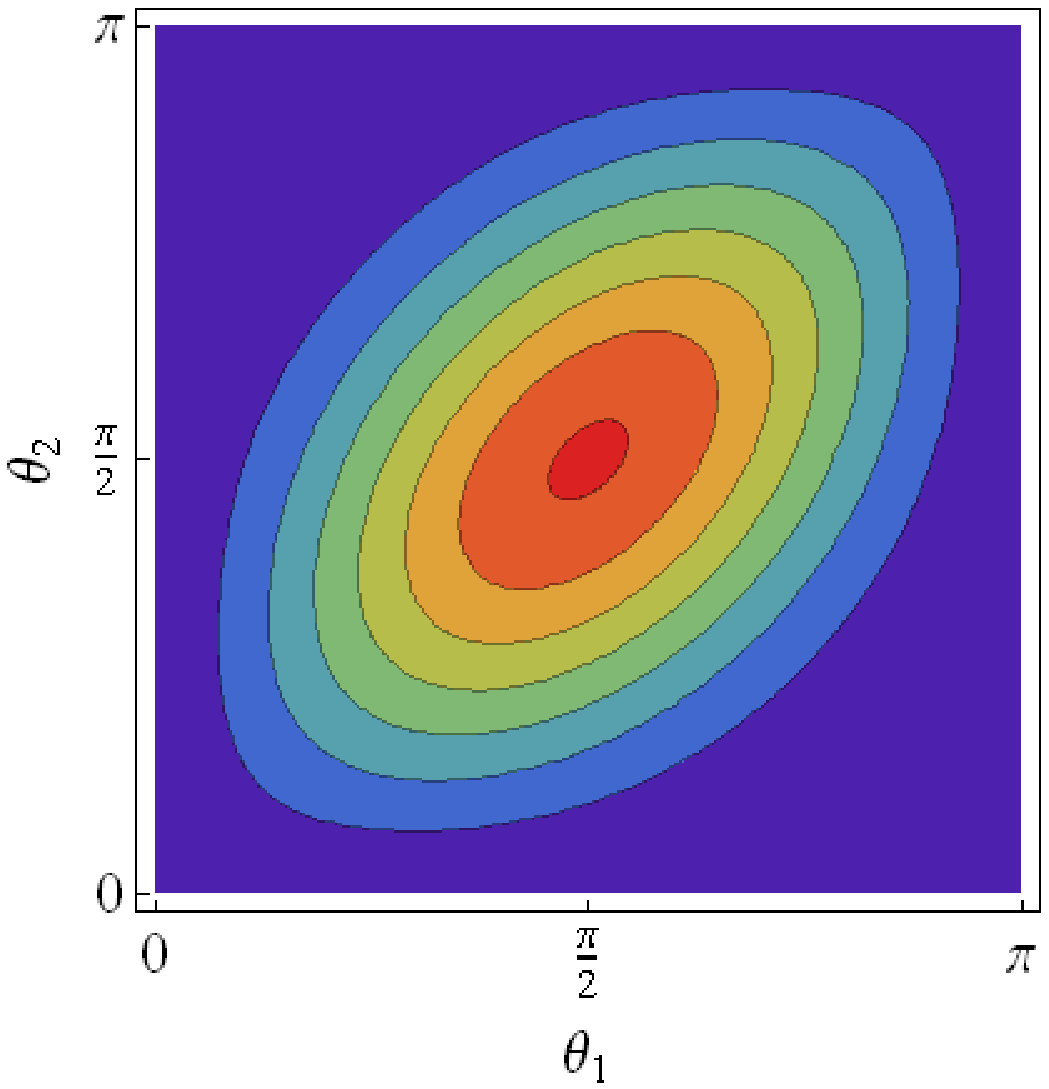}
(a2)
\end{minipage}\hspace{0.0cm}
\begin{minipage}[t]{7.3cm}
\centering
\includegraphics[width=\textwidth]{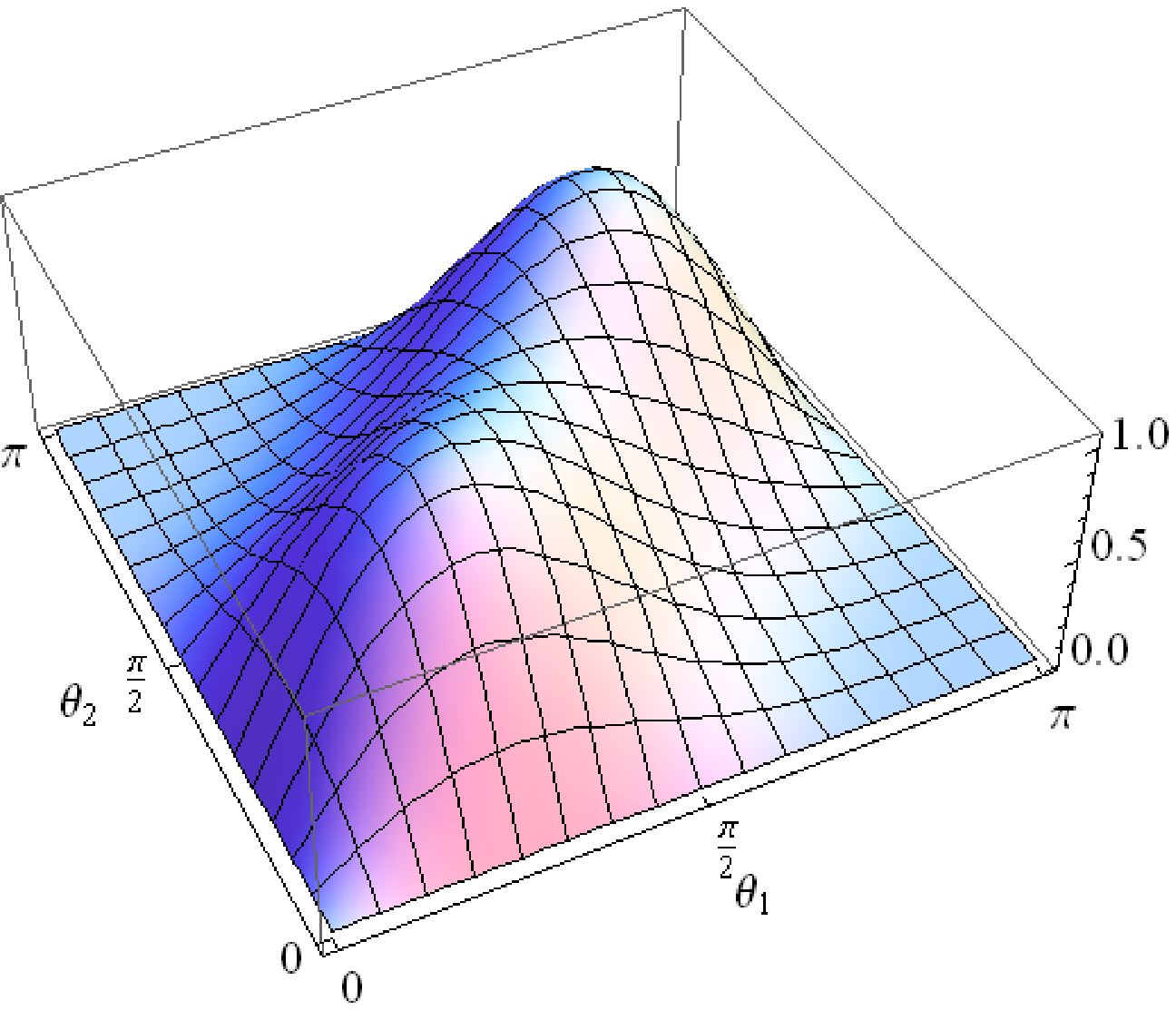}
(b1)
\end{minipage}\hspace{1.0cm}
\begin{minipage}[t]{6.7cm}
\centering
\includegraphics[width=\textwidth]{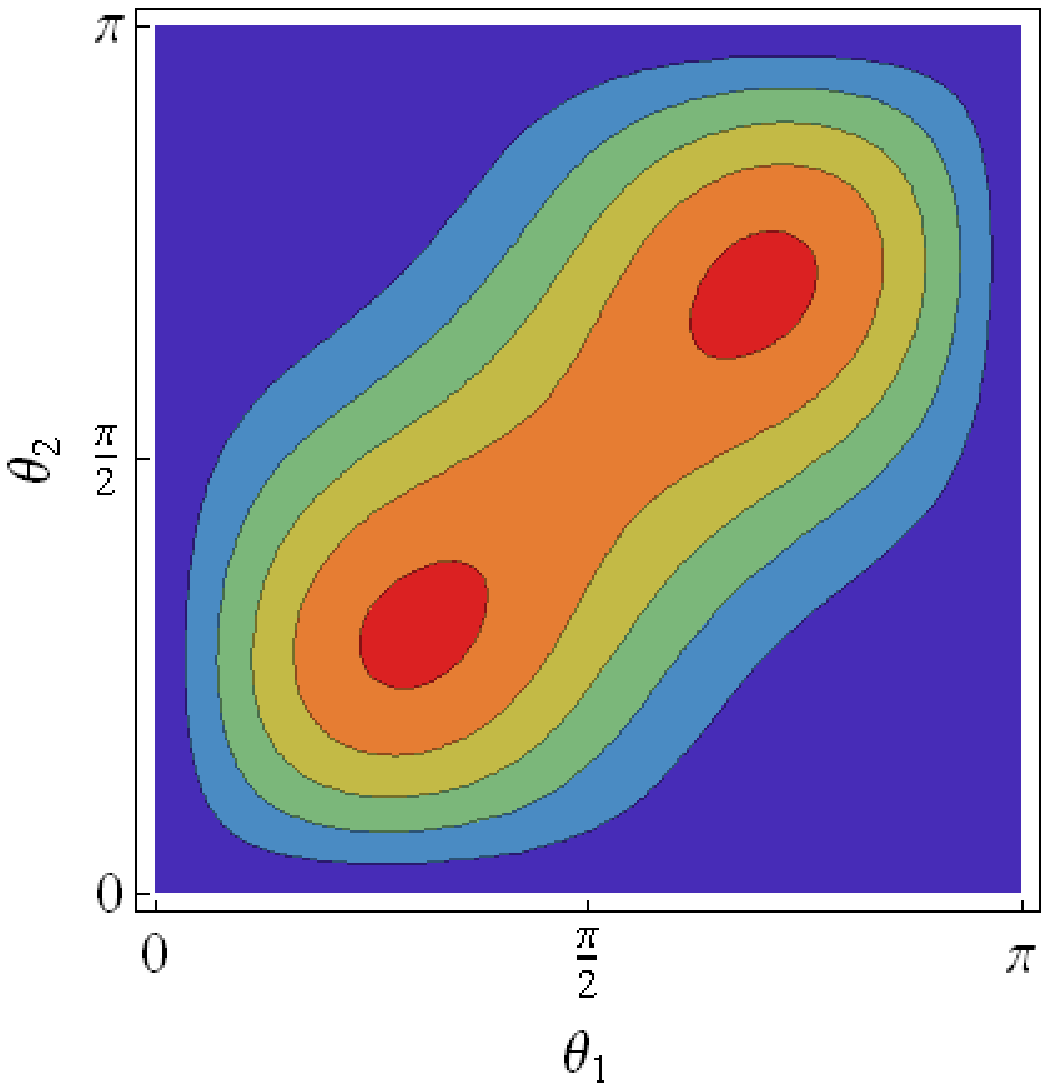}
(b2)
\end{minipage}\hspace{0.0cm}
\begin{minipage}[t]{7.3cm}
\centering
\includegraphics[width=\textwidth]{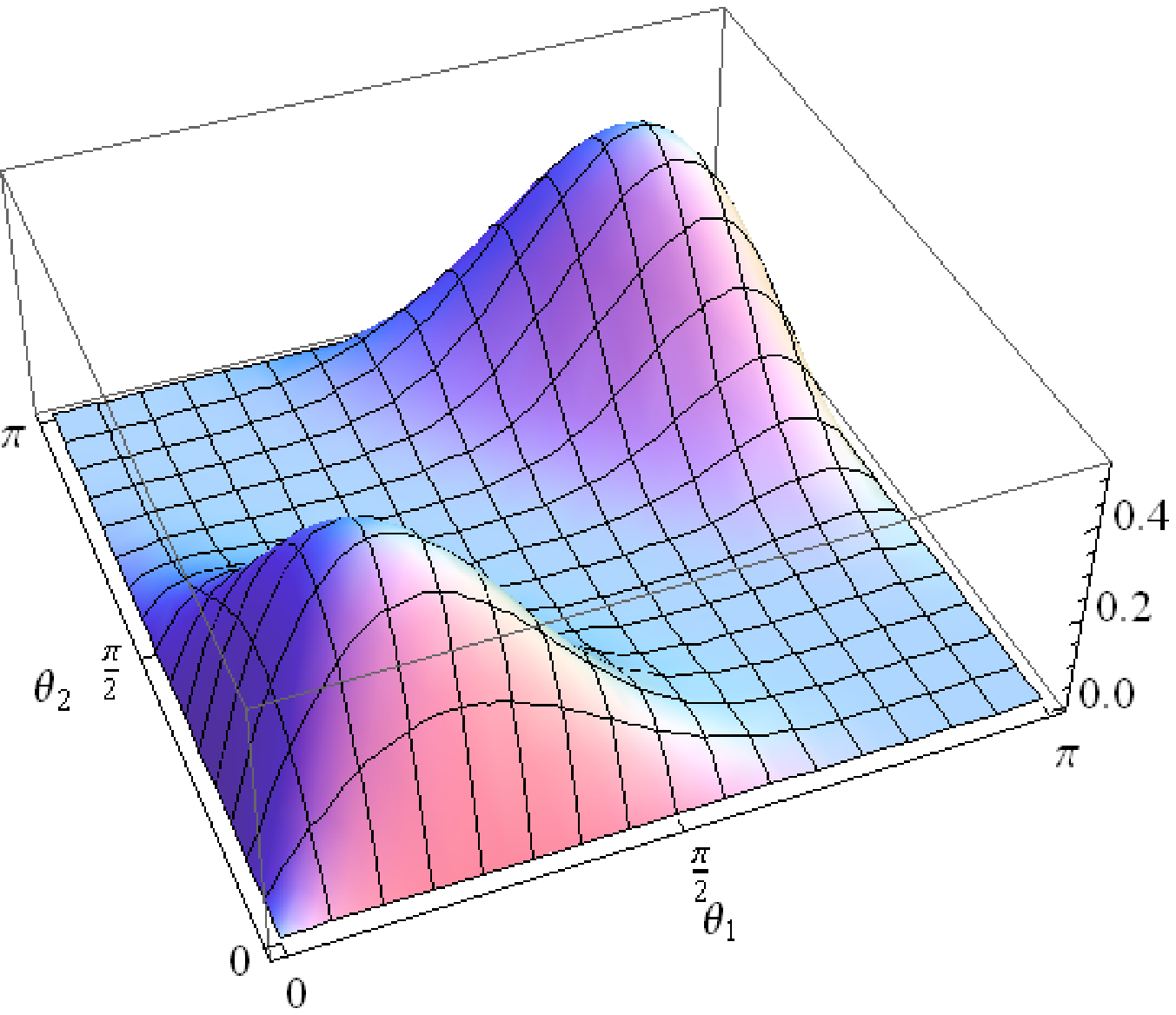}
(c1)
\end{minipage}\hspace{1.0cm}
\begin{minipage}[t]{6.7cm}
\centering
\includegraphics[width=\textwidth]{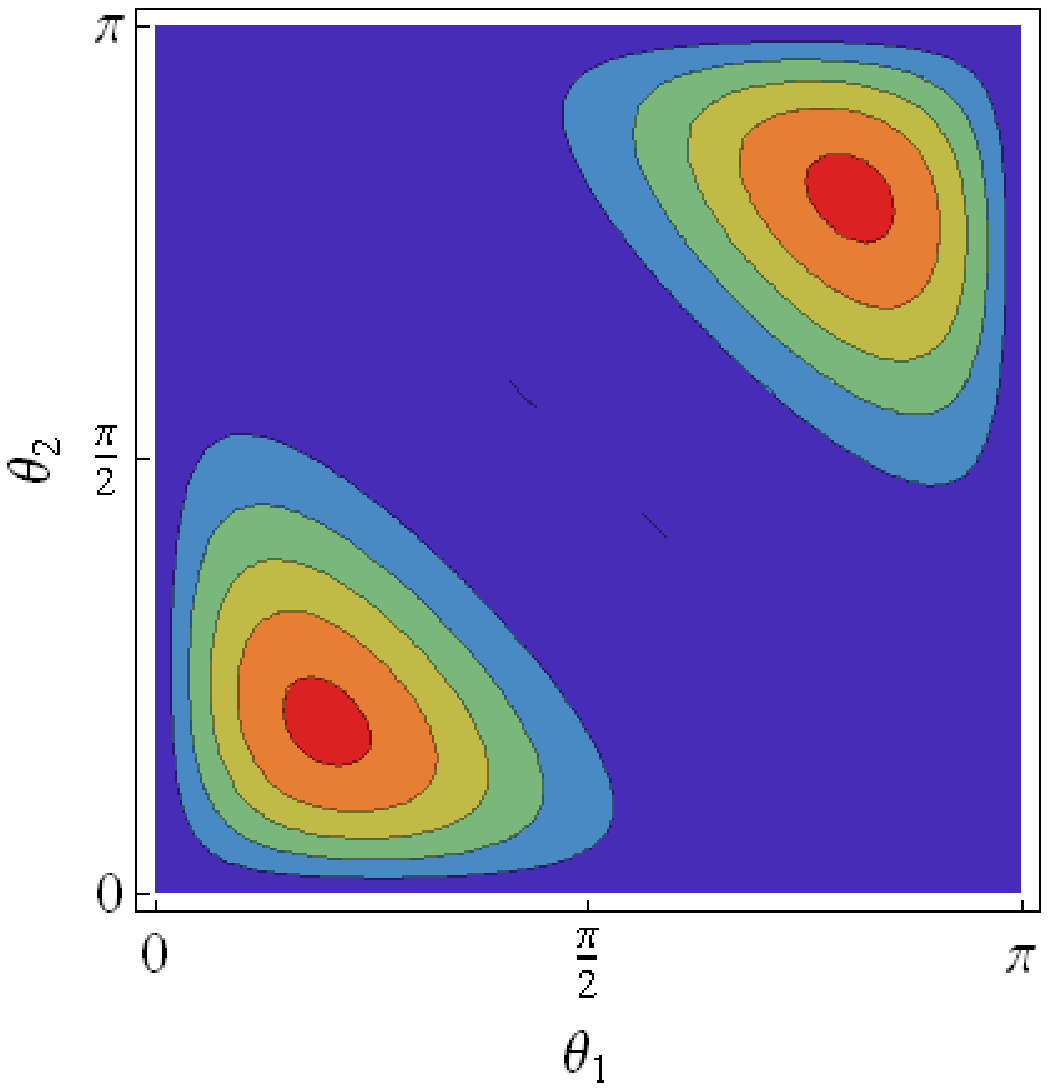}
(c2)
\end{minipage}\hspace{0.0cm}
\parbox{0.95\textwidth}{\caption{
\it 
The angular distributions for decays of $X_0 \to W^- W^+ \to \ell^- \ell^+ {\bar \nu} \nu$
given by (\protect\ref{J20a5J00}) $\times \sin{\theta_1}\, \sin{\theta_2}$ for (a) $\phi =0$,
(b) $\phi ={\pi}/2$, and (c) $\phi =\pi$.
\label{fig:WWJ00phi0}}}
\end{figure}

\begin{figure}
\centering
\begin{minipage}[t]{10.0cm}
\centering
\includegraphics[width=\textwidth]{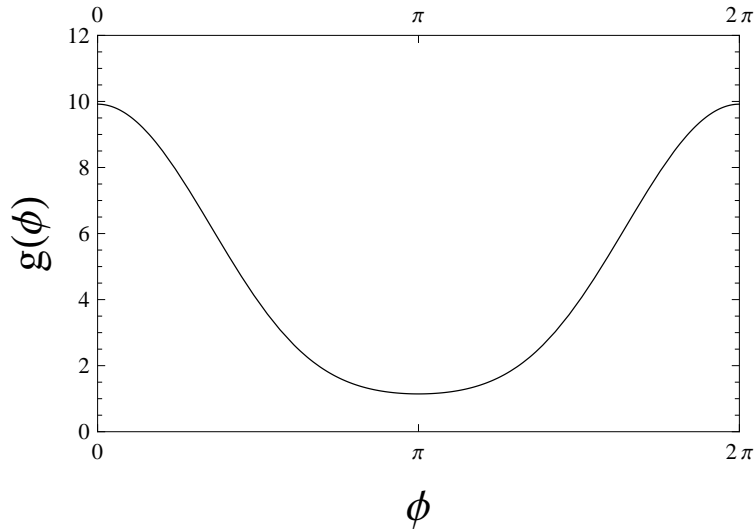}
\end{minipage}\hspace{0.0cm}
\vspace*{1.0cm}
\parbox{0.95\textwidth}{\caption{
\it
The azimuthal angular distribution $g(\phi)$ given in (\ref{J20a6J00})
for $X_0 \to W^- W^+ \to \ell^- \ell^+ {\bar \nu} \nu$ decay.
\label{dsigmadphi_spin0.eps}}} 
\end{figure}

For comparison, we now review the case of a spin-zero boson $X_0$.
From the Clebsch-Gordan coefficients in
\begin{equation}
|00 \rangle\ =\ {\sqrt{1\over 3}}|1+1\rangle|1-1\rangle-{\sqrt{1\over 3}}|10\rangle|10\rangle
+{\sqrt{1\over 3}}|1-1\rangle|1+1\rangle\ ,
\label{cg3w0}
\end{equation}
at the vertex $X_0 \to W^- W^+$ for the $|JJ^z\rangle=|00\rangle$ state of $X_0$,
we see that the polarizations of the $W^-$ and $W^+$ are in the following coherent state:
\begin{equation}
{\sqrt{1\over 3}}\, \epsilon_1^+\epsilon_2^-\ -\ {\sqrt{1\over 3}}\, \epsilon_1^0\epsilon_2^0\ +\
{\sqrt{1\over 3}}\, \epsilon_1^-\epsilon_2^+\ .
\label{J20ww0}
\end{equation}
The state of the $W^- W^+$ pair produced at the
$X_0 \to W^- W^+$ vertex of the spin-zero Higgs particle given by (\ref{J20ww0}) 
is proportional to $\eta_{\alpha\beta}$.

Then, from (\ref{a4w}, \ref{a5w}, \ref{a6w}), (\ref{a4wlp}, \ref{a5wlp}, \ref{a6wlp}) and (\ref{J20ww0}),
we see that the amplitude ${\cal M}$ of (\ref{aab1w1lmlp}) is given by the following coherent amplitude
(omitting a factor $1/(2 \sqrt{2}p)^2$):
\begin{equation}
a\, e^{+i\phi}\ +\ be^{-i\phi}\ +\ c\ ,
\label{J20a1J00}
\end{equation}
where $\phi\equiv \phi_1 -\phi_2$ and
\begin{eqnarray}
a&=&-\ {\sqrt{1\over 3}}\ (1-\cos{\theta_1})\ (1-\cos{\theta_2})
\label{J20a2J00}\\
b&=&-\ {\sqrt{1\over 3}}\ (1+\cos{\theta_1})\ (1+\cos{\theta_2})
\label{J20a3J00}\\
c&=&-\ {\sqrt{1\over 3}}\ \ 2\ \sin{\theta_1}\ \sin{\theta_2} \ .
\label{J20a4J00}
\end{eqnarray}
The absolute square of (\ref{J20a1J00}) is given by
\begin{equation}
\Big(\ a\, e^{-i\phi}\ +\ be^{i\phi}\ +\ c\ \Big)\,
\Big(\ a\, e^{i\phi}\ +\ be^{-i\phi}\ +\ c\ \Big)\ =\
a^2+b^2+c^2+2(a+b)c\cos{\phi}+2ab\cos{2\phi} \ .
\label{J20a5J00}
\end{equation}
In Fig. \ref{fig:WWJ00phi0}
we plot (\ref{J20a5J00}) $\times \sin{\theta_1}\, \sin{\theta_2}$
(which is proportional to ${d^2\sigma / d\theta_1d\theta_2}$)
for $\phi =0$, $\pi/2$ or ${3\pi/2}$, and $\pi$.
The azimuthal angle distribution resulting from the integration
\begin{equation}
g(\phi)\ \equiv \
\int_0^\pi \sin{\theta_1}\, d{\theta_1}\, \int_0^\pi \sin{\theta_2}\, d{\theta_2}\
\Big[ \ {\rm Eq. (\ref{J20a5J00})} \ \Big]
\ ,
\label{J20a6J00}
\end{equation}
is presented in Fig.~\ref{dsigmadphi_spin0.eps}.

Comparing the results presented in Fig. \ref{dsigmadphi_spin0.eps} for the $X_0$ case with
the fact that $g(\phi)$ is constant for the $X_2$ case as shown in the previous subsection,
we see their clear difference in the angular correlations between the $\ell^\pm$. 
This suggests that a careful study
of the $\ell^\pm$ angular distributions might offer some discrimination between the
spin-two and spin-zero hypotheses. We note that the ATLAS~\cite{ATLASH} and CMS~\cite{CMSH} $W^+ W^-$
event selections are based on the $\ell^\pm$ angular distributions predicted in the spin-zero
case~\cite{DD}, see the angular distributions for the data, backgrounds and a possible
$H \to W^- W^+ \to \ell^- \ell^+ {\bar \nu} \nu$ signal in~\cite{ATLASWW,CMSWW}, 
and are likely to have reduced efficiencies for the spin-two case. 
However, any conclusions on the possible hypotheses would
require realistic simulations of the $W^- W^+ \to \ell^- \ell^+ {\bar \nu} \nu$ final states in an LHC detector.


\subsection{Dilepton invariant mass distributions}

\begin{figure}
\centering
\vspace*{1.5cm}
\begin{minipage}[t]{12.5cm}
\centering
\includegraphics[width=\textwidth]{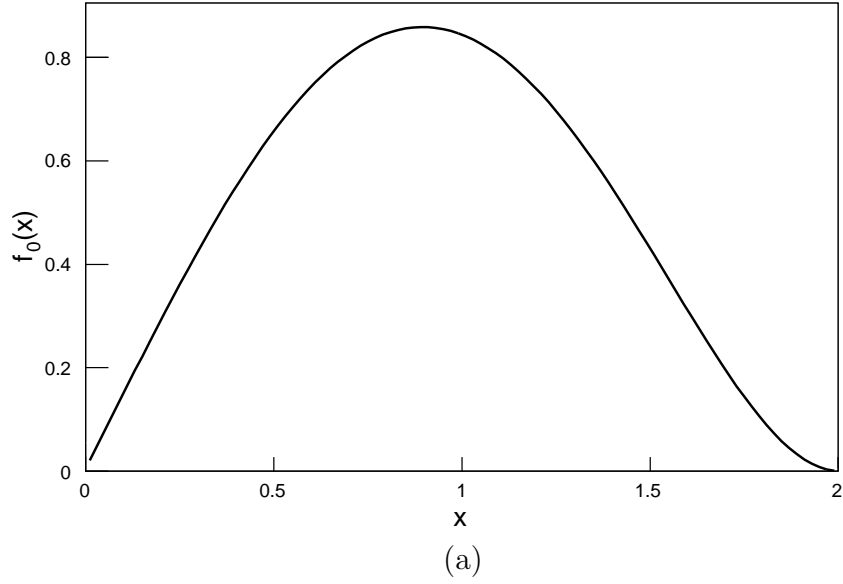}
(a)
\end{minipage}\hspace{1.0cm}
\vspace{0.5cm}
\begin{minipage}[t]{12.5cm}
\centering
\includegraphics[width=\textwidth]{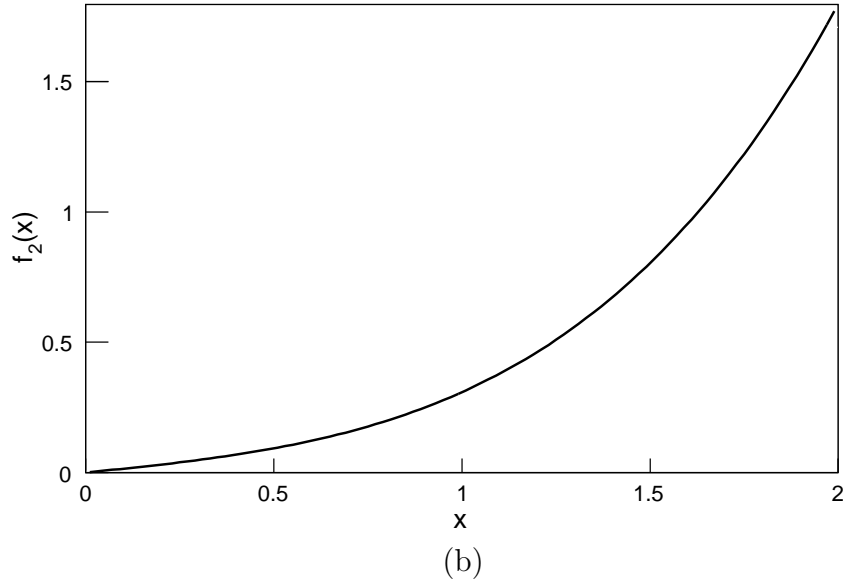}
(b)
\end{minipage}\hspace{1.0cm}
\vspace{1.5cm}
\parbox{0.95\textwidth}{\caption{
\it
Distributions in the $\ell^- \ell^+$ invariant mass, $m_{ll}$, for (a) the spin-zero case $X_0$ and (b) for the spin-two case $X_2$.
The plotted quantities are $f_{0,2}(x) \equiv {1\over \sigma}{d\sigma \over dx}$ where $x \equiv 2 m_{ll}/m_W$, so that
$x=2$ corresponds to $m_{ll}=m_W={m\over 2}$, where $m$ is the mass of $X_0$ or $X_2$.
\label{fig:mass}}}
\end{figure}

We conclude this Section by displaying in Fig.~\ref{fig:mass} the distributions in the $\ell^- \ell^+$ invariant mass, $m_{ll}$,
for $X_0 \to W^- W^+ \to \ell^- \ell^+ {\bar \nu} \nu$ (a) and
$X_2 \to W^- W^+ \to \ell^- \ell^+ {\bar \nu} \nu$ (b). As could be expected from
the differences in the angular distributions discussed above, and on the basis of helicity
arguments, the $\ell^- \ell^+$ invariant mass distribution peaks at a larger value in the
$X_2$ case than in the $X_0$ case. This offers, in principle, another way to discriminate
between the two possible spin assignments. We note that~\cite{ATLASWW,CMSWW} also compare
data for $m_{ll}$ with simulations of $X_0 \to W^- W^+ \to \ell^- \ell^+ {\bar \nu} \nu$ and experimental backgrounds.

\section{Summary}

We have presented in this paper analyses of the angular distributions
that could be expected in the $\gamma \gamma$
and $W^- W^+$ decays of a hypothetical spin-two state $X_2$ produced at the
LHC via gluon-gluon collisions, assuming that its couplings coincide with those
expected for a massive Kaluza-Klein graviton. Under this hypothesis, such a spin-two particle 
would be produced in a definite combination of polarization states, and the
polar angle distribution of the $\gamma \gamma$ final state would be
predictable and non-isotropic in the $X_2$ rest frame, and hence
distinguishable in principle from the isotropic $\gamma \gamma$ decays
of a hypothetical spin-zero boson $X_0$. Likewise, the angular
correlations between the $\ell^\pm$ produced in $X_2 \to W^- W^+$
decays are predictable and distinct from those in $X_0 \to W^- W^+$
decays. In this paper we have analyzed the case where the $W^- W^+$ pair
is at rest, which may be a suitable first approximation to the case of
a state with mass $\sim 125$~GeV decaying into $W W^*$.

This analytical study would require detailed simulations for either
ATLAS and/or CMS before one could conclude whether, in practice,
these angular distributions could be used to provide supplementary
information about the spin of the hypothetical particle that may be
responsible for the excesses of events seen at $\sim 125$~GeV
by both ATLAS and CMS. We think that the effort of making such 
simulations should be worthwhile, in view of the excesses of
$\gamma \gamma$ and $W^- W^+$ events already seen. 

As a preliminary step in this direction, we have initiated a project
to simulate off-shell effects and the CMS and ATLAS experimental event selections,
detection efficiencies and acceptances using {\tt PYTHIA} and
{\tt Delphes}. Preliminary results of these simulations indicate that angular distributions
discussed here do not vary substantially for candidate Higgs masses between 165
and 125 GeV, and continue to offer good discrimination between the spin-zero and
spin-two hypotheses. Fuller details will be published elsewhere~\cite{EFHSY}.

Many
other approaches to analyzing the possible spin of a Higgs candidate
rely on lepton angular correlations in $ZZ \to 4 \ell^\pm$
final states~\cite{2,4,7,9,9a,10,11,12,13}.
These would provide considerably more information, but are
limited by statistics likely to be available in the foreseeable future.

\subsubsection*{Acknowledgements}

The work of J.E. is supported partly by the London
Centre for Terauniverse Studies (LCTS), using funding from the European
Research Council 
via the Advanced Investigator Grant 267352.
The work of D.S.H. is supported partly by
Korea Foundation for International Cooperation of Science \& Technology (KICOS)
and National Research Foundation of Korea (2011-0005226).
D.S.H. thanks CERN for its hospitality while working on this subject.

\end{document}